\documentclass[%
 aip,
 amsmath,amssymb,%
 reprint,%
]{revtex4-1}

\usepackage{graphicx}% Include figure files
\usepackage{dcolumn}% Align table columns on decimal point
\usepackage{bm}% bold math
\usepackage[utf8]{inputenc}
\usepackage[T1]{fontenc}
\usepackage{mathptmx}
\usepackage{etoolbox}
\usepackage{url}
\usepackage{subcaption}
\usepackage[english]{babel}
\makeatletter
\def\@email#1#2{%
 \endgroup
 \patchcmd{\titleblock@produce}
  {\frontmatter@RRAPformat}
  {\frontmatter@RRAPformat{\produce@RRAP{*#1\href{mailto:#2}{#2}}}\frontmatter@RRAPformat}
  {}{}
}%
\makeatother
\begin{document}

\preprint{AIP/123-QED}

\title[Order--Disorder Transitions and Thermal Pathways in Frustrated 2D Colloidal Crystals]{Order--Disorder Transitions and Thermal Pathways in Frustrated 2D Colloidal Crystals}
% Force line breaks with \\

\author{Alexandre Vargas}
 \affiliation{Programa de Pós-Graduação em Física, Instituto de Física e Matemática, Universidade Federal de Pelotas. Caixa Postal 354, CEP 96001-970, Pelotas, RS, Brazil}%Lines break automatically or can be forced with \\

\author{Thiago Puccinelli}
\affiliation{Instituto de F\'isica, Universidade de S\~ao Paulo (USP), 05508-090, S\~ao Paulo, Brasil}

\author{José Rafael Bordin}
 \affiliation{Departamento de Física, Instituto de Física e Matemática, Universidade Federal de Pelotas. Caixa Postal 354, CEP 96001-970, Pelotas, RS, Brazil}%Lines break automatically or can be forced with \\
\email{jrbordin@ufpel.edu.br}
\date{\today}% It is always \today, today,

\begin{abstract}
We employ extensive $NPT$ molecular dynamics simulations to explore the thermal transitions of two-dimensional colloidal crystals interacting via a core-softened potential with competing length scales. The system stabilizes three distinct solid phases, namely low-density triangular (LDT), stripe, and kagome, which exhibit markedly different responses to heating and cooling. Our simulations reveal that the LDT and kagome phases melt via first-order transitions, but only the former recrystallizes smoothly. The kagome phase displays strong hysteresis and metastability, while the stripe phase undergoes a continuous and nearly reversible transformation. These results highlight the role of lattice geometry and frustration in shaping non-universal melting and freezing pathways in 2D soft matter.
\end{abstract}

\maketitle

\section{Introduction}

The thermal response of two-dimensional (2D) colloidal systems remains a rich and complex area of study, particularly when interparticle interactions promote structural competition and frustration. In such systems, phase transitions can deviate significantly from classical expectations, exhibiting non-universal melting and freezing pathways governed by subtle differences in lattice geometry, interaction range, and symmetry constraints \cite{Mermin1968, Kosterlitz1973, Halperin1978, Nelson1979, Strandburg1988, Bernard2011, Kosterlitz2016}.

Core-softened (CS) interactions represent a versatile class of models that capture such phenomena, characterized by a hard core followed by a soft repulsive shoulder \cite{deOliveira2006, deOliveira2006b, Likos2001, Likos06}. CS potentials are effective in modeling colloids functionalized with polymeric or dendritic coronas and give rise to competing length scales that stabilize unconventional crystalline phases—such as stripe, honeycomb, and kagome lattices \cite{jagla1998, Jagla1999a, Camp2006, Pattabhiraman2017, Jain2014, Ryzhov2023, Fomin2019b, OlsonReichhardt2010}. These systems are known to exhibit reentrant melting, cluster formation, multiple solid polymorphisms, and metastable states across 2D and 3D cases \cite{Prestipino2012, Fomin2021, Bordin2018, Wassermair2024, Kapfer2015}.

Despite extensive literature on melting in CS and related 2D systems \cite{Chen1995, Jaster1999, Marcus1996, Nosenko2009, Guillamn2009, Horn2013, Deutschlander2013, Hua2024, MassanaCid2024}, the freezing process—and especially the comparison between melting and freezing—remains much less explored \cite{Strandburg1988, Bernard2011, Deutschlander2013, Karthika2016}. This gap is especially significant in frustrated or polymorphic contexts, where ordering and disordering may follow fundamentally different kinetic routes \cite{OlsonReichhardt2010, Cunuder2019, Libal2018, Nowack2019}. In such settings, melting cannot be assumed to mirror freezing, and simultaneous investigation of both directions is critical for comprehensive understanding \cite{Anderson2017, Charbonneau2024, Royall2024}.

Recent theoretical and experimental advances—such as two-step melting studies \cite{Bernard2011, Charbonneau2024} and disorder-induced transitions \cite{Deutschlander2013, Horn2013}—still typically focus on heating protocols. Few works systematically compare heating and cooling trajectories, even though such dual-path methodological approaches are essential to uncover metastability, hysteresis, and reversibility \cite{Tsiok2015, MendozaCoto2024, Kapfer2015, Ryzhov2023}. These gaps highlight the need for studies that treat melting and freezing on equal footing \cite{Stoycheva2000, OlsonReichhardt2010, Nowack2019, Fomin2021, MendozaCoto2024}.

In this work, we provide a dual-path investigation of melting and freezing in a 2D colloidal system governed by a core-softened potential that stabilizes three distinct crystalline phases: low-density triangular (LDT), stripe, and kagome \cite{Cardoso2021, Nogueira2022, Nogueira2023}. Rather than centering on applicability of KTHNY or hexatic frameworks alone \cite{Mermin1968, Kosterlitz1973, Halperin1978, Nelson1979, Kapfer2015, Hua2024, Puccinelli25}, we focus on the influence of lattice geometry, symmetry and frustration in shaping each thermal path. Our emphasis is on reversibility, hysteresis, kinetic traps, and partial ordering across both melting and recrystallization. The results reveal that LDT and kagome phases both melt via first-order transitions, but only LDT recrystallizes smoothly; the stripe phase undergoes continuous transitions in both directions, whereas kagome remains trapped in metastable states \cite{Cardoso2021, Nogueira2022, OlsonReichhardt2010, Nowack2019, Cunuder2019, Libal2018}. 

These findings indicates that melting and freezing in 2D soft matter are inherently non-universal and strongly shaped by symmetry, interaction competition, and thermodynamic path dependency. A dual-pathway approach is therefore necessary to fully characterize phase behavior in low-dimensional frustrated systems, including functional colloids, thin films, and two-dimensional molecular assemblies \cite{Menath2021, Feller2022, Camerin2022, Royall2024, Russo2018}.

The remainder of this paper is organized as follows: Section~\ref{sec:model} introduces the interaction model and simulation setup. Section~\ref{sec:results} presents our thermodynamic and structural analysis of the melting and freezing transitions. Finally, Section~\ref{sec:conclusion} summarizes the main conclusions and discusses future research directions.

\section{Interaction Model and Simulation Details}
\label{sec:model}

\begin{figure}[h!]
\centering
\includegraphics[width=6cm]{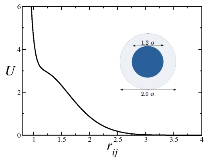}
\caption{Effective core-softened interaction potential used in this work. The potential consists of a Lennard-Jones term and a Gaussian repulsion centered at $r_0 = 0.7\sigma$. Inset: illustration of the two relevant length scales—short-range core ($r_1 \approx 1.2\sigma$) and soft shell ($r_2 \approx 2.0\sigma$).}
\label{fig-potencial}
\end{figure}

We consider a two-dimensional system composed of $N = 5000$ particles interacting via a core-softened (CS) potential originally proposed by de Oliveira et al.~\cite{deOliveira2006}. The interaction consists of a Lennard-Jones-like repulsion–attraction term combined with a Gaussian repulsive shoulder, resulting in two characteristic length scales:

\begin{equation}
U_{\text{CS}}(r) = 4\epsilon \left[ \left(\frac{\sigma}{r}\right)^{12} - \left(\frac{\sigma}{r}\right)^6 \right] + u_0 \exp\left[-\frac{1}{c_0^2} \left(\frac{r - r_0}{\sigma} \right)^2 \right],
\label{eq:CS}
\end{equation}

\noindent where $u_0 = 5\epsilon$, $c_0^2 = 1.0$, and $r_0 = 0.7\sigma$. This form introduces a repulsive shoulder at intermediate distances, resulting in a potential with a stiff core around $r_1 \approx 1.2\sigma$ and a soft shell near $r_2 \approx 2.0\sigma$—as shown in Fig.\ref{fig-potencial}. The competition between these two scales drives polymorphism and frustration in the system\cite{deOliveira2006b, BarrosdeOliveira2010}.

Molecular Dynamics simulations were performed using the isothermal-isobaric ($NpT$) ensemble, implemented in the LAMMPS package~\cite{lammps}. The $NpT$ ensemble was chosen to allow the system volume to fluctuate in response to temperature changes. This is essential once volume constraints could distort the underlying lattice symmetries.

We investigated three reduced pressures: $p=1.0$, $p=4.0$, and $p=9.0$, which stabilize the low-density triangular (LDT), stripe, and kagome phases, respectively~\cite{Cardoso2021, Nogueira2022}. For each isobar, we performed heating and cooling cycles between temperatures $T=0.10$ and $T=0.20$, using steps of $\Delta T = 0.02$. Each temperature point involved $1 \times 10^6$ equilibration steps, followed by $5 \times 10^7$ production steps. The final configuration of each step was used to initialize the next, allowing for quasi-adiabatic evolution.

The temperature and pressure were controlled via Nosé–Hoover thermostat and barostat, with damping constants $Q_T = 0.1$ and $Q_p = 1.0$, respectively. The time step was set to $\delta t = 0.01$, and averages and snapshots were recorded every $10^4$ steps.

To characterize structural transitions, we calculated both short-range and long-range orientational order parameters $\Psi_l$ and $G_l(\mathbf{r})$, defined via Voronoi-based neighbor analysis and evaluated with the \texttt{freud} Python library~\cite{freud2020}. Translational ordering was quantified via the radial distribution function $g(r)$ and its integral $\tau$\cite{Errington2001}. We also computed isobaric susceptibilities $\chi_{\Psi_l}$ and $\chi_\tau$ to capture order-parameter fluctuations. Finally, the specific heat at constant pressure, $C_p$, was estimated both via numerical differentiation of the enthalpy and from enthalpy fluctuations~\cite{allen2017}.

\section{Results and Discussion}
\label{sec:results}

\begin{figure}[h!]
\centering
\includegraphics[width=8cm]{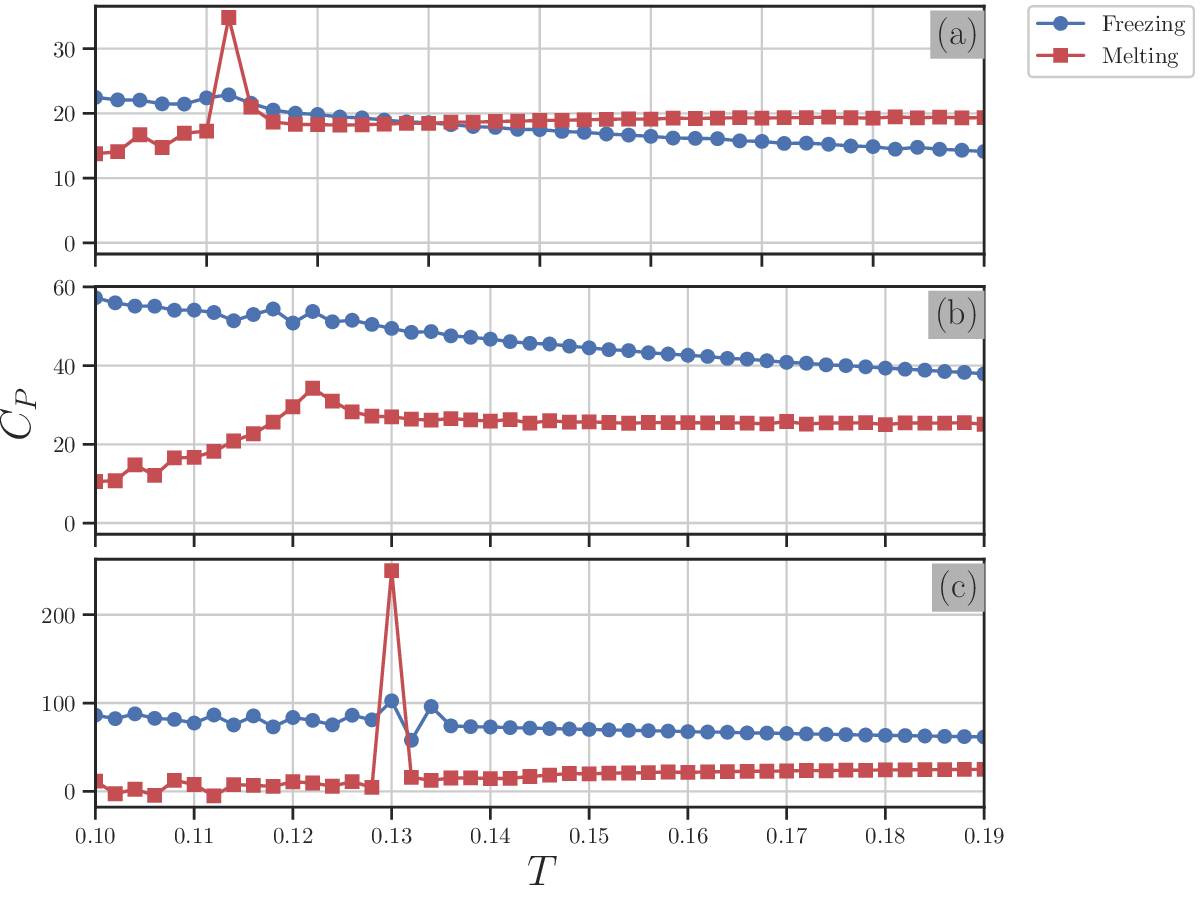}
\caption{Specific heat at constant pressure $C_p$ as a function of temperature for (a) the LDT phase at $p = 1.0$, (b) the stripe phase at $p = 4.0$, and (c) the kagome phase at $p = 9.0$. Red curves correspond to the melting path and blue curves to the freezing path. Discontinuities and hysteresis indicate different mechanisms for melting and crystallization.}
\label{figcp}
\end{figure}

We now present a comparative analysis of the melting and freezing behavior of the three polymorphic phases stabilized by the core-softened potential: the low-density triangular (LDT), stripe, and kagome structures. Our focus is on identifying the thermodynamic signatures, structural transformations, and the degree of thermal hysteresis observed along each thermal cycle, following recent efforts to characterize kinetic asymmetries in 2D colloidal systems with frustration and multiple competing orderings.

Figure~\ref{figcp} shows the temperature dependence of the isobaric specific heat $C_p$ for each phase. In the LDT phase (Figure~\ref{figcp}a), we observe a clear jump in $C_p$ during melting, indicating a first-order transition characterized by the abrupt loss of both translational and orientational order, as expected for crystalline structures with high symmetry and compact packing. This discontinuity contrasts with the smoother, rounded maximum found upon freezing, suggestive of classical nucleation-and-growth dynamics and the progressive elimination of metastable fluid clusters. The pronounced thermal hysteresis between the heating and cooling paths is a hallmark of first-order transitions, reflecting the presence of energy barriers and kinetic bottlenecks associated with defect annihilation and recrystallization.

\begin{figure*}[ht]
    \centering
    \begin{subfigure}[b]{0.47\textwidth}
        \centering
        \includegraphics[width=7.5cm]{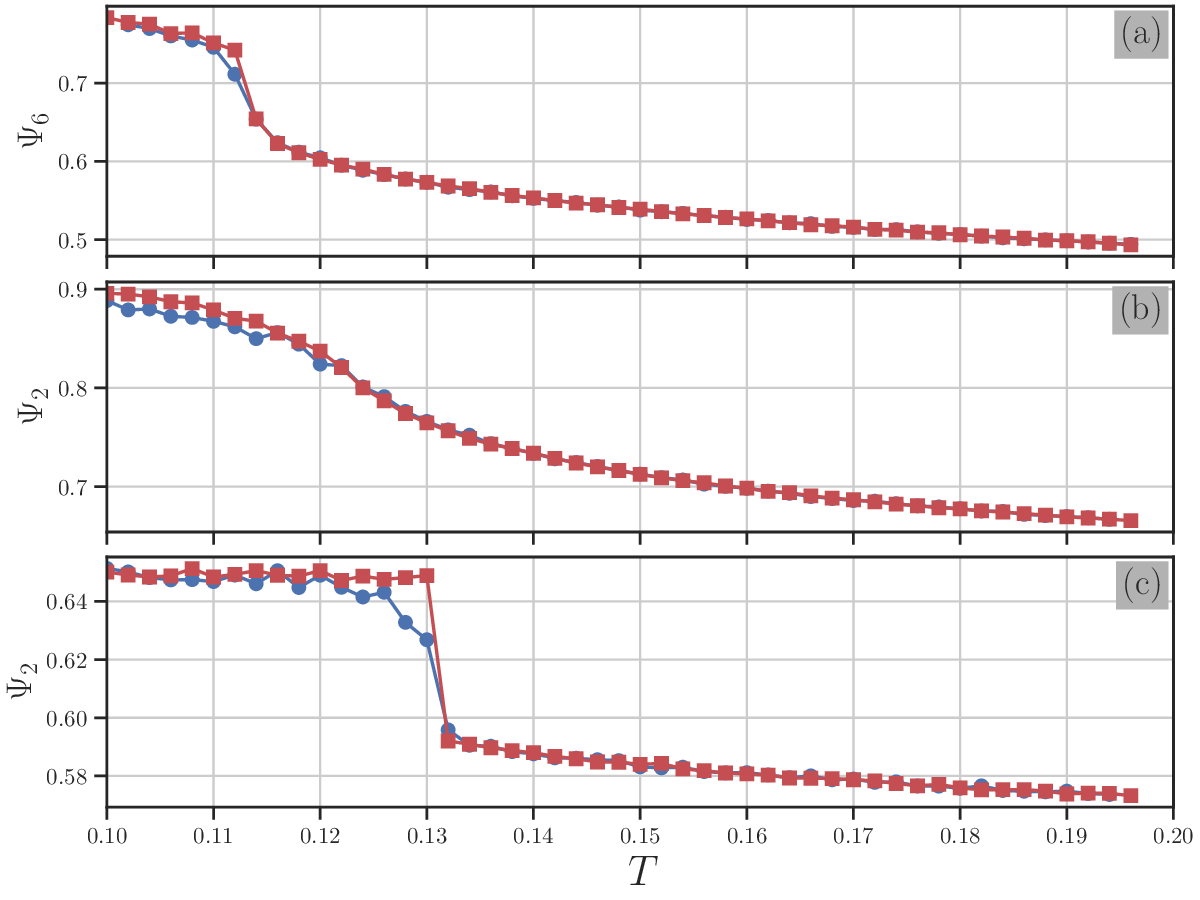}
        \caption{Orientational order parameter \(\Psi_l\)}
    \end{subfigure}
    \hfill
    \begin{subfigure}[b]{0.47\textwidth}
        \centering
        \includegraphics[width=8cm]{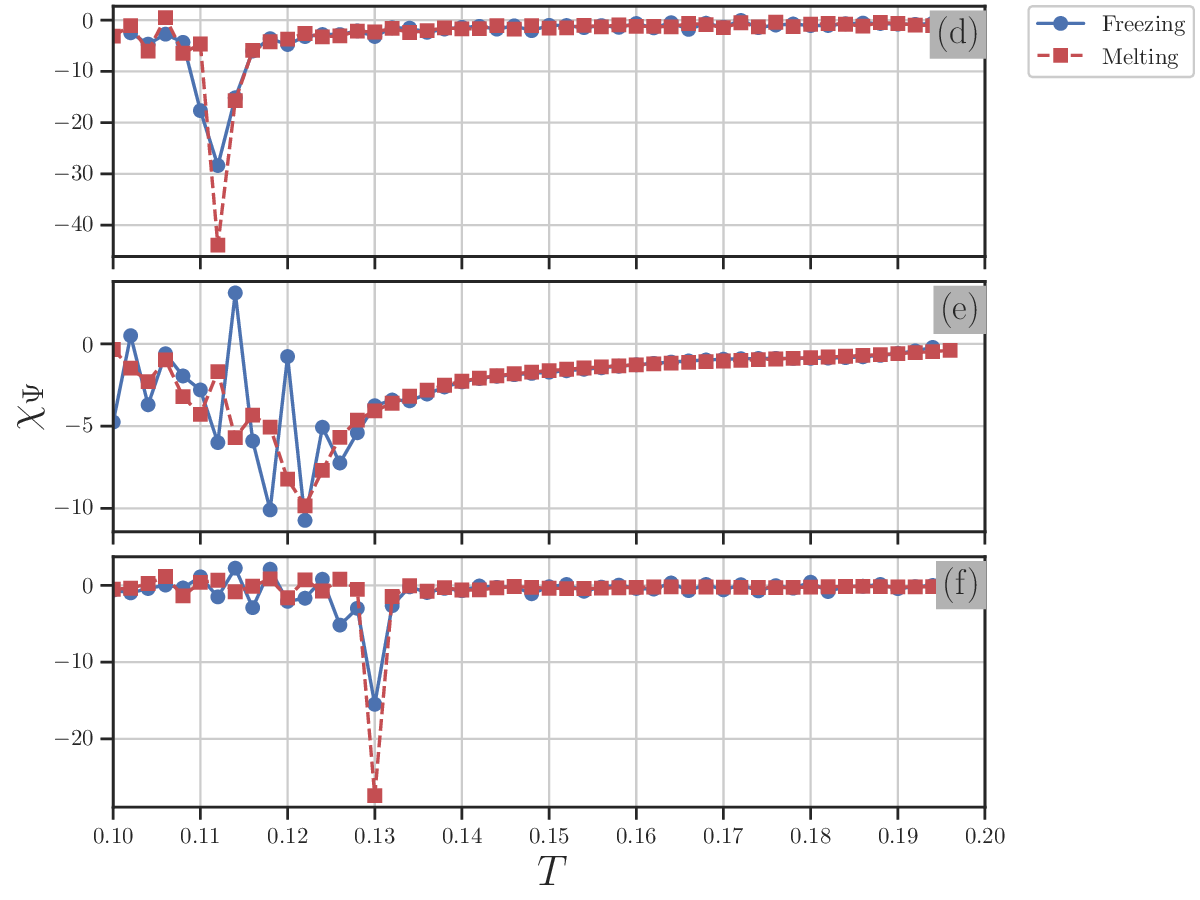}
        \caption{Isobaric orientational susceptibility \(\chi_{\Psi_l}\)}
    \end{subfigure}
    \caption{Orientational order parameter and its susceptibility as functions of temperature for: (a,d) LDT with \(l=6\), (b,e) stripe with \(l=2\), and (c,f) kagome with \(l=2\). Red and blue lines correspond to heating and cooling paths, respectively. The symmetry index \(l\) is chosen to match each crystal's dominant local structure.}
    \label{fig_psis}
\end{figure*}

In the stripe phase (Figure~\ref{figcp}b), $C_p$ exhibits a broad, continuous peak during heating, consistent with a second-order-like transition and the absence of latent . Upon cooling, the specific heat does not exhibit a sharp feature, suggesting minimal hysteresis and a reversible transition. This behavior reflects the quasi-one-dimensional symmetry and internal flexibility of the stripe configuration~\cite{Pattabhiraman2017}, which allow for gradual rearrangement and defect unbinding during both melting and freezing~\cite{OlsonReichhardt2010}.

The kagome phase displays the strongest asymmetry between melting and freezing. As shown in Figure~\ref{figcp}c, $C_p$ presents a sharp discontinuity during heating, consistent with a first-order transition. In contrast, the freezing path is marked by irregular fluctuations in $C_p$, indicative of metastability and frustrated crystallization. These oscillations are compatible with the presence of intermediate amorphous or clustered states~\cite{Nowack2019} that delay the formation of a fully ordered kagome lattice. The resulting thermal hysteresis further supports the classification of the transition as first-order and underscores the intrinsic geometrical frustration and complex local connectivity of the kagome network.

These results also underscore the critical role played by the thermodynamic path: the system's response during heating differs significantly from that during cooling, revealing a marked hysteresis in phases with first-order transitions~\cite{Kapfer2015, Bernard2011, Tsiok2015}. This asymmetry reflects not only differences in equilibrium stability, but also the influence of kinetic barriers and metastable intermediates that are accessible (or avoided) depending on the direction of the thermal cycle~\cite{Strandburg1988, Libal2018}. Thus, any complete characterization of phase transitions in such systems must account for both melting and freezing trajectories, as each path selectively probes different portions of the energy landscape~\cite{Charbonneau2024, Royall2024}.

To gain further insight into the structural transformations during melting and freezing, we analyze the orientational order parameter $\Psi_l$ and its corresponding isobaric susceptibility $\chi_{\Psi_l}$, as shown in Figure~\ref{fig_psis}. These observables allow us to track the degree of local angular ordering and its sensitivity to thermal fluctuations across the transition.

In the LDT phase, which possesses sixfold bond-orientational symmetry, $\Psi_6$ remains close to unity at low temperatures, reflecting a well-ordered triangular lattice. Upon heating, $\Psi_6$ drops abruptly at the melting point, signaling the rapid breakdown of angular correlations across the system. This sharp loss of orientational order is accompanied by a pronounced peak in the susceptibility $\chi_{\Psi_6}$, reinforcing the identification of a first-order transition. Notably, the freezing trajectory shows a smoother but still well-defined recovery of $\Psi_6$, consistent with a classical nucleation-and-growth process in which crystalline domains progressively reemerge and coarsen~\cite{jagla1998}. The thermal hysteresis observed between the heating and cooling paths is a hallmark of first-order transitions, reflecting energy barriers and the influence of interfacial defects in the reorganization process. This asymmetry also underscores the kinetic differences between melting and crystallization in this phase~\cite{Kapfer2015}.

In the stripe phase, the orientational order is better captured by $\Psi_2$, which quantifies the degree of alignment between parallel lanes of particles. Here, $\Psi_2$ decays smoothly with increasing temperature, indicating a gradual loss of stripe alignment. The corresponding susceptibility $\chi_{\Psi_2}$ exhibits a broad, rounded maximum centered at the melting point, with no discontinuity. Upon cooling, both $\Psi_2$ and $\chi_{\Psi_2}$ retrace their paths with minimal deviation, pointing to a highly reversible transformation. This behavior is compatible with a continuous transition driven by defect unbinding, where stripe deformation and reconnection proceed without abrupt symmetry breaking~\cite{malescio2003}. The internal flexibility of the stripe phase facilitates this soft response, enabling the formation of elongated, entangled structures that gradually dissolve into a polymer-like fluid during melting. Conversely, during freezing, aligned lanes reform continuously from the fluid, minimizing the energetic cost of reorganization~\cite{jagla1998}.

In contrast, the kagome phase exhibits markedly different behavior. The orientational order parameter $\Psi_2$, which here captures the relative orientation of the tri-triangular motifs forming the kagome pattern, drops discontinuously during heating and recovers sharply upon cooling. The susceptibility $\chi_{\Psi_2}$ shows narrow, intense peaks along both trajectories, confirming the first-order character of the transition. However, in comparison to the LDT phase, the freezing path is significantly more irregular. The recovery of $\Psi_2$ during cooling is accompanied by intermittent fluctuations and temporary plateaus, indicating that the system becomes trapped in metastable or partially ordered configurations before fully regaining the kagome lattice.

This behavior is intimately connected to the geometric frustration inherent to the kagome network~\cite{han2008}. The complex local topology requires the simultaneous arrangement of multiple three-fold motifs, making the formation of large, defect-free domains kinetically challenging. As a result, small imperfections or misaligned clusters can prevent complete ordering, leading to slow crystallization dynamics and persistent structural noise. This frustration-induced metastability is reflected not only in $\chi_{\Psi_2}$, but also in the specific heat $C_p$ and other observables, and distinguishes the kagome freezing process from that of the more symmetric LDT phase.

\begin{figure*}[ht]
    \centering
    \begin{subfigure}[b]{0.47\textwidth}
        \centering
        \includegraphics[width=7.5cm]{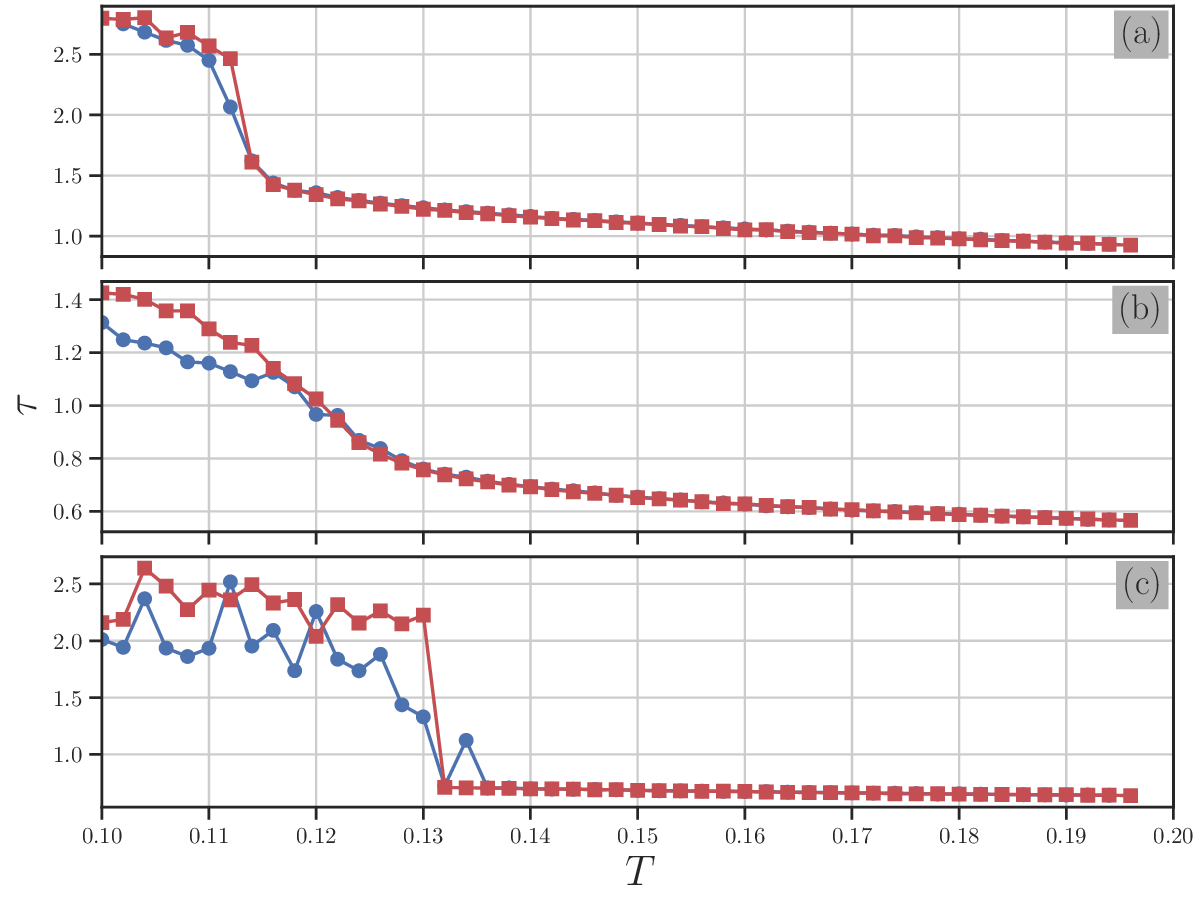}
        \caption{Translational order parameter \(\tau\)}
    \end{subfigure}
    \hfill
    \begin{subfigure}[b]{0.47\textwidth}
        \centering
        \includegraphics[width=8cm]{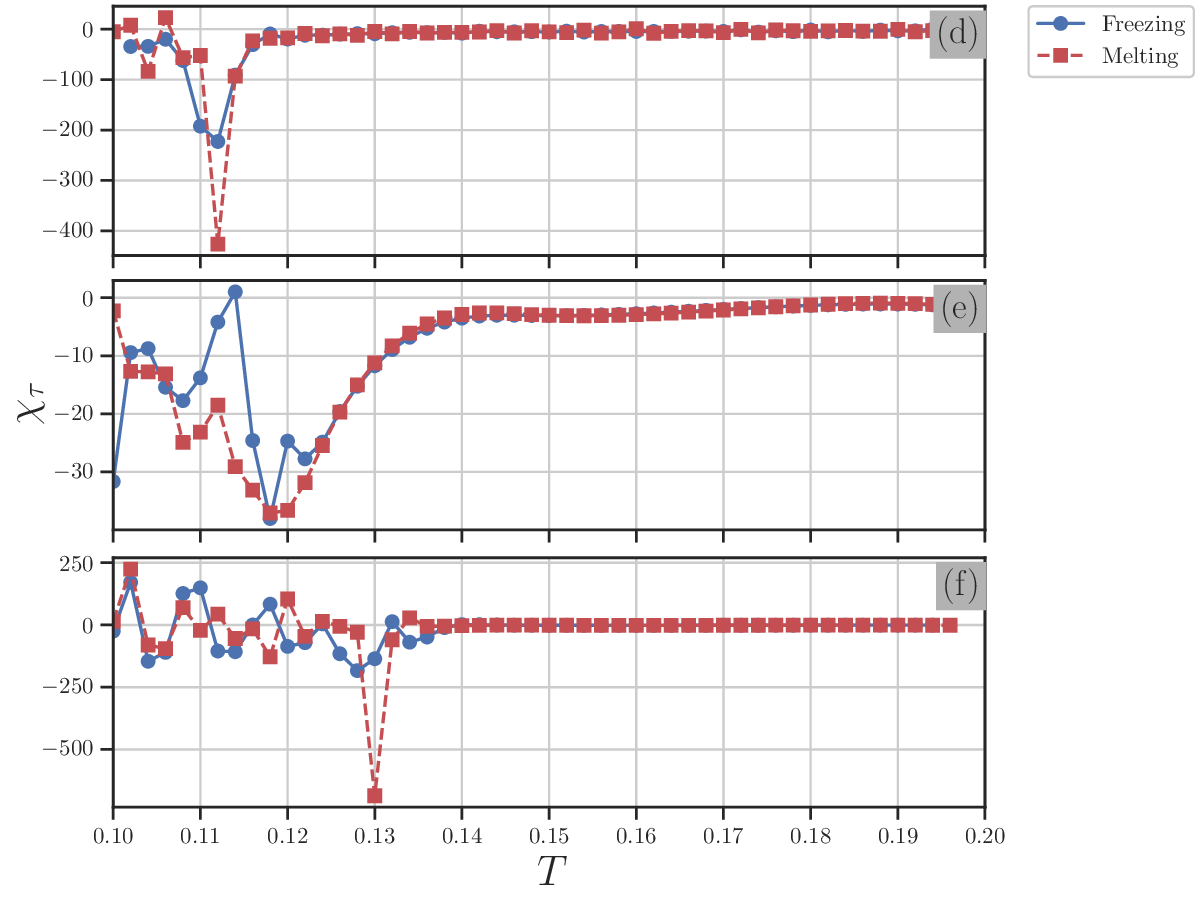}
        \caption{Isobaric translational susceptibility \(\chi_\tau\)}
    \end{subfigure}
    \caption{Translational order and susceptibility as functions of temperature for the three studied phases: (a,d) LDT, (b,e) stripe, and (c,f) kagome. Red and blue lines represent melting and freezing trajectories, respectively.}
    \label{fig_tau}
\end{figure*}

The translational order parameter $\tau$, shown in Figure~\ref{fig_tau}, provides complementary information to the orientational observables, quantifying how density modulations decay with distance and thereby revealing the degree of long-range positional ordering in the system. This measure is particularly sensitive to the collapse or emergence of crystalline periodicity during phase transitions.

In the LDT phase, $\tau$ maintains high values at low temperatures, reflecting the regular spacing of particles in the triangular lattice. As temperature increases, a sharp drop in $\tau$ is observed at the melting point, indicating a sudden loss of translational symmetry. This abrupt decay parallels the behavior of the orientational order parameter $\Psi_6$, confirming that melting in this phase is strongly first-order, involving a simultaneous breakdown of both angular and positional order. Upon cooling, $\tau$ recovers rapidly but not perfectly symmetrically, again pointing to the presence of nucleation barriers and a characteristic hysteresis loop associated with the energy cost of forming ordered domains from a disordered fluid. This partial asymmetry in the freezing curve is consistent with a scenario of classical crystallization, in which localized regions of order grow and coalesce through interface-limited dynamics.

For the stripe phase, $\tau$ behaves in a fundamentally different manner. Instead of an abrupt jump, it decreases gradually with temperature, exhibiting a broad, rounded susceptibility peak centered around the melting point. This behavior supports a continuous or weakly first-order transition scenario. The stripe configuration, being quasi-one-dimensional, does not rely on global translational symmetry to the same extent as the LDT or kagome phases. Instead, positional order is confined along individual lanes, and melting involves a progressive undulation and eventual disintegration of these aligned chains. As the stripes bend, reconnect, and lose coherence, the system smoothly transitions into a fluid state composed of elongated, polymer-like clusters with residual memory of the stripe alignment. The reversibility observed in the cooling path of $\tau$, which retraces the heating curve with minimal hysteresis, further supports this interpretation. Freezing proceeds via the gradual realignment and consolidation of these clusters into parallel lanes, bypassing the need for long-range positional coherence at early stages.

In the kagome phase, the behavior of $\tau$ again reflects a strongly first-order transition. As in the LDT case, $\tau$ drops steeply during melting, revealing a rapid disruption of positional order. However, the recovery upon cooling is far less regular: $\tau$ increases in a stepwise, irregular fashion, suggesting that translational order reforms only intermittently and is often disrupted by competing local motifs. This erratic behavior mirrors that seen in the orientational order parameter $\Psi_2$, and together they point to the presence of metastable intermediates and frustrated crystallization pathways. The complex unit cell of the kagome lattice, which requires precise placement of particles at specific geometric angles, renders the establishment of coherent long-range order particularly susceptible to kinetic traps. Even small angular misalignments can prevent the full development of translational symmetry, leading to a fragmented and slow ordering process. This interpretation is supported by the presence of oscillations in $C_p$ and $\chi_{\Psi_2}$ during cooling, all indicative of a highly nontrivial energy landscape with multiple local minima.

\begin{figure*}[ht]
    \centering
    \begin{subfigure}[b]{0.45\textwidth}
        \centering
        \includegraphics[width=6cm,height=6cm]{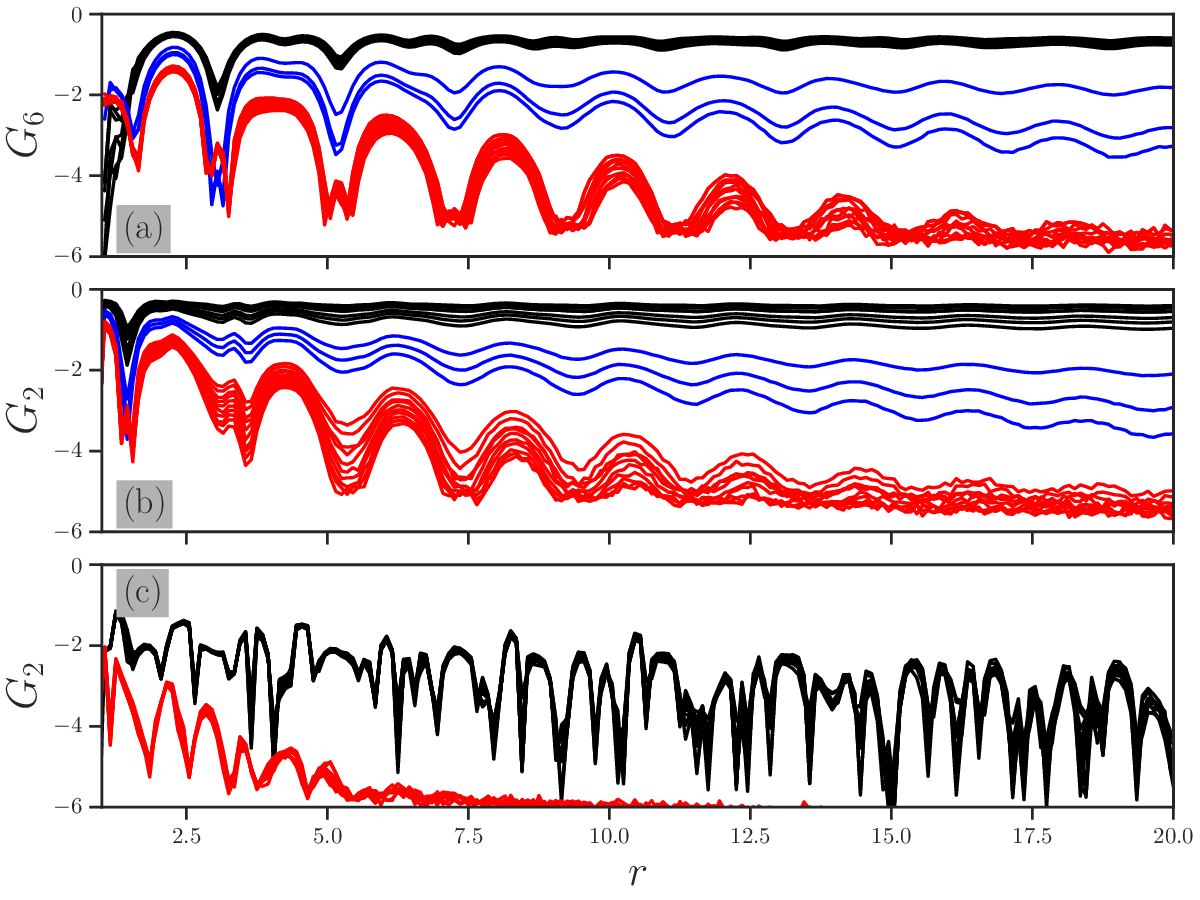}
        \caption{Melting}
    \end{subfigure}
    \hfill
    \begin{subfigure}[b]{0.50\textwidth}
        \centering
        \includegraphics[width=6cm]{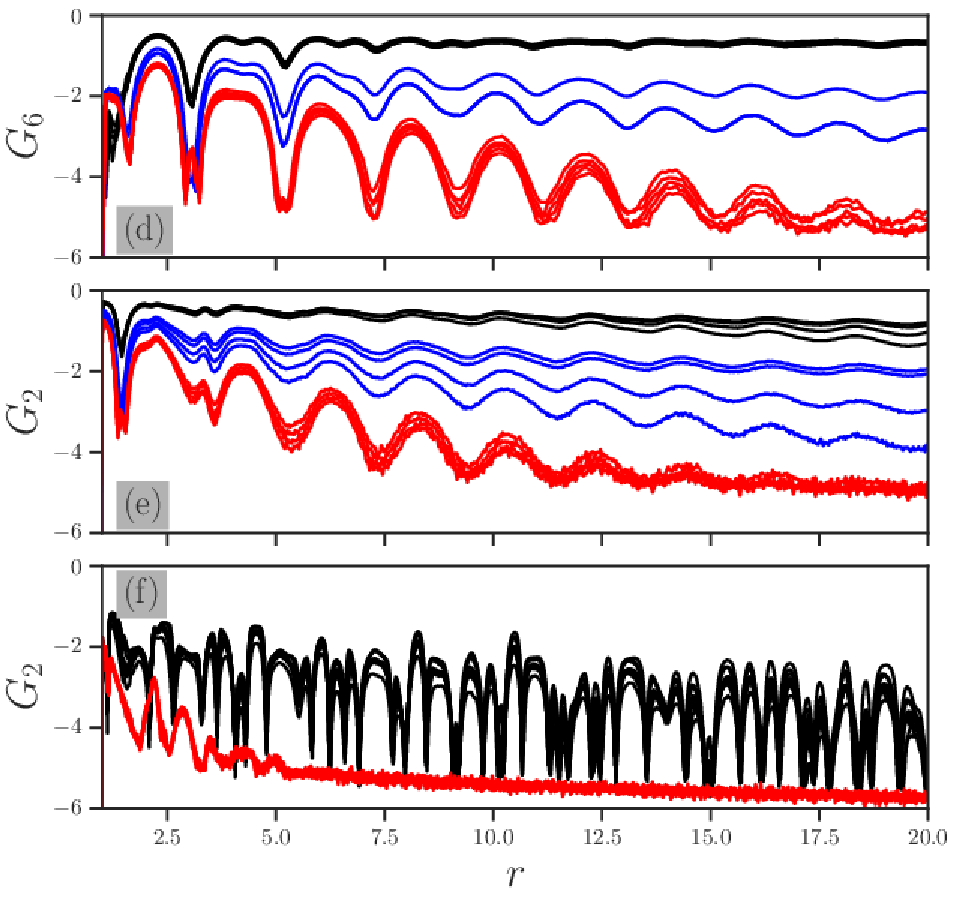}
        \caption{Freezing}
    \end{subfigure}
\caption{Bond-orientational correlation functions \( G_l(r) \) during (a) melting and (b) freezing for the three studied phases: top — LDT with \( G_6(r) \), middle — stripe with \( G_2(r) \), bottom — kagome with \( G_2(r) \). Black curves correspond to the crystalline phase, blue curves to temperatures in the fluid phase near the phase transition, and red curves to higher temperatures in the disordered regime.}
    \label{fig-gs}
\end{figure*}

To assess the persistence of orientational order across melting and freezing cycles, we analyze the spatial decay of the bond-orientational correlation functions $G_l(r)$, shown in Figure~\ref{fig-gs}. In the LDT and stripe phases, $G_l(r)$ retains significant angular coherence over long distances in the crystalline regime. During melting, this coherence weakens progressively, suggesting a gradual loss of structural alignment. Notably, the decay remains slow even near the transition, indicating that these phases preserve a degree of orientational memory over extended spatial ranges. However, it is important to emphasize that, although slow decays of $G_l(r)$ are observed near melting in the LDT and stripe phases, the data do not clearly support the emergence of a distinct hexatic regime. This reinforces the interpretation of non-universal melting pathways, governed not by a universal two-step mechanism, but by frustration, lattice symmetry, and kinetic constraints specific to each phase.

In the LDT phase, this progressive decay of $G_6(r)$ is followed by a partial recovery upon cooling, with long-range correlations reestablished relatively quickly—consistent with a well-defined nucleation pathway and the formation of compact crystalline domains. The stripe phase behaves similarly: $G_2(r)$ maintains coherence throughout the thermal cycle. This suggests that the quasi-one-dimensional geometry of the stripe network facilitates reversible rearrangements without complete orientational loss, in agreement with the minimal hysteresis observed in $C_p$, $\Psi_2$, and $\tau$.

In contrast, the kagome phase exhibits a sharp collapse of orientational correlations. The function $G_2(r)$ decays rapidly beyond a few interparticle distances during melting and shows no significant recovery during freezing. This behavior reflects the spontaneous breakdown of the local tri-triangular motifs and the difficulty of reconstructing the kagome pattern due to topological constraints and geometrical frustration. The lack of orientational coherence during freezing is consistent with the strong metastability and irregular thermal signatures previously observed.

A comparative overview of the three phases reveals how symmetry, frustration, and the chosen thermodynamic path jointly govern the nature and reversibility of the transitions. The LDT phase exhibits a clear first-order melting with sharp discontinuities in all order parameters and $C_p$, followed by a smoother but complete recrystallization upon cooling. In contrast, the kagome phase also melts abruptly but fails to recover fully during freezing, reflecting its intrinsic geometric frustration and complex local constraints. The stripe phase, by comparison, undergoes continuous and nearly symmetric transitions, with smooth variations in both structural and thermodynamic observables.

These findings highlight the critical role of the thermodynamic trajectory in revealing hidden asymmetries and metastable behaviors that would remain inaccessible from a single-direction analysis. The hysteresis loops, the degree of reversibility, and the presence (or absence) of intermediate states all depend strongly on the direction of the thermal path. This underlines the necessity of dual-pathway studies to uncover the full spectrum of phase-transition mechanisms in 2D soft matter systems with competing interactions.

\section{Conclusions and Perspectives}
\label{sec:conclusion}

We investigated the melting and freezing behavior of a two-dimensional colloidal system governed by a core-softened interaction potential that stabilizes three distinct crystalline phases: a low-density triangular (LDT) lattice, a stripe phase, and a kagome structure. Our molecular dynamics simulations reveal that each phase exhibits a characteristic thermal response: both the LDT and kagome structures melt via first-order transitions, but only the LDT phase recrystallizes in a regular and continuous manner. The kagome lattice, by contrast, shows strong hysteresis and metastability upon cooling, consistent with the influence of geometric frustration. The stripe phase follows a continuous and nearly reversible pathway in both directions, facilitated by its quasi-one-dimensional geometry.

These results suggest that melting and freezing in two-dimensional soft matter are inherently non-universal, shaped by lattice symmetry, geometric frustration, and the presence of competing interaction scales. By analyzing complete thermal cycles, our work highlights how asymmetries between disordering and reordering processes emerge, with structural reversibility serving as a key signature to distinguish polymorphic pathways. Beyond characterizing specific crystalline phases, our findings provide a framework to understand how geometry and thermal trajectory govern phase behavior in low-dimensional systems. This dual-path perspective offers a guidance for the design and control of crystallization dynamics in functional colloidal monolayers, thin films, and other emerging 2D soft materials.

\section*{Funding Declaration}

J.R.B. acknowledges financial support from CNPq (grant numbers 405479/2023-9, 441728/2023-5, and 304958/2022-0A). A.V.I. acknowledges support from CAPES (Finance Code 001). Thiago Puccinelli acknowledges support from FAPESP.

\bibliography{bib}

\begin{thebibliography}{58}%
\makeatletter
\providecommand \@ifxundefined [1]{%
 \@ifx{#1\undefined}
}%
\providecommand \@ifnum [1]{%
 \ifnum #1\expandafter \@firstoftwo
 \else \expandafter \@secondoftwo
 \fi
}%
\providecommand \@ifx [1]{%
 \ifx #1\expandafter \@firstoftwo
 \else \expandafter \@secondoftwo
 \fi
}%
\providecommand \natexlab [1]{#1}%
\providecommand \enquote  [1]{``#1''}%
\providecommand \bibnamefont  [1]{#1}%
\providecommand \bibfnamefont [1]{#1}%
\providecommand \citenamefont [1]{#1}%
\providecommand \href@noop [0]{\@secondoftwo}%
\providecommand \href [0]{\begingroup \@sanitize@url \@href}%
\providecommand \@href[1]{\@@startlink{#1}\@@href}%
\providecommand \@@href[1]{\endgroup#1\@@endlink}%
\providecommand \@sanitize@url [0]{\catcode `\\12\catcode `\$12\catcode `\&12\catcode `\#12\catcode `\^12\catcode `\_12\catcode `\%12\relax}%
\providecommand \@@startlink[1]{}%
\providecommand \@@endlink[0]{}%
\providecommand \url  [0]{\begingroup\@sanitize@url \@url }%
\providecommand \@url [1]{\endgroup\@href {#1}{\urlprefix }}%
\providecommand \urlprefix  [0]{URL }%
\providecommand \Eprint [0]{\href }%
\providecommand \doibase [0]{http://dx.doi.org/}%
\providecommand \selectlanguage [0]{\@gobble}%
\providecommand \bibinfo  [0]{\@secondoftwo}%
\providecommand \bibfield  [0]{\@secondoftwo}%
\providecommand \translation [1]{[#1]}%
\providecommand \BibitemOpen [0]{}%
\providecommand \bibitemStop [0]{}%
\providecommand \bibitemNoStop [0]{.\EOS\space}%
\providecommand \EOS [0]{\spacefactor3000\relax}%
\providecommand \BibitemShut  [1]{\csname bibitem#1\endcsname}%
\let\auto@bib@innerbib\@empty
%</preamble>
\bibitem [{\citenamefont {Mermin}(1968)}]{Mermin1968}%
  \BibitemOpen
  \bibfield  {author} {\bibinfo {author} {\bibfnamefont {N.~D.}\ \bibnamefont {Mermin}},\ }\bibfield  {title} {\enquote {\bibinfo {title} {Crystalline order in two dimensions},}\ }\href {\doibase 10.1103/physrev.176.250} {\bibfield  {journal} {\bibinfo  {journal} {Physical Review}\ }\textbf {\bibinfo {volume} {176}},\ \bibinfo {pages} {250–254} (\bibinfo {year} {1968})}\BibitemShut {NoStop}%
\bibitem [{\citenamefont {Kosterlitz}\ and\ \citenamefont {Thouless}(1973)}]{Kosterlitz1973}%
  \BibitemOpen
  \bibfield  {author} {\bibinfo {author} {\bibfnamefont {J.~M.}\ \bibnamefont {Kosterlitz}}\ and\ \bibinfo {author} {\bibfnamefont {D.~J.}\ \bibnamefont {Thouless}},\ }\bibfield  {title} {\enquote {\bibinfo {title} {Ordering, metastability and phase transitions in two-dimensional systems},}\ }\href {\doibase 10.1088/0022-3719/6/7/010} {\bibfield  {journal} {\bibinfo  {journal} {Journal of Physics C: Solid State Physics}\ }\textbf {\bibinfo {volume} {6}},\ \bibinfo {pages} {1181–1203} (\bibinfo {year} {1973})}\BibitemShut {NoStop}%
\bibitem [{\citenamefont {Halperin}\ and\ \citenamefont {Nelson}(1978)}]{Halperin1978}%
  \BibitemOpen
  \bibfield  {author} {\bibinfo {author} {\bibfnamefont {B.~I.}\ \bibnamefont {Halperin}}\ and\ \bibinfo {author} {\bibfnamefont {D.~R.}\ \bibnamefont {Nelson}},\ }\bibfield  {title} {\enquote {\bibinfo {title} {Theory of two-dimensional melting},}\ }\href {\doibase 10.1103/physrevlett.41.121} {\bibfield  {journal} {\bibinfo  {journal} {Physical Review Letters}\ }\textbf {\bibinfo {volume} {41}},\ \bibinfo {pages} {121–124} (\bibinfo {year} {1978})}\BibitemShut {NoStop}%
\bibitem [{\citenamefont {Nelson}\ and\ \citenamefont {Halperin}(1979)}]{Nelson1979}%
  \BibitemOpen
  \bibfield  {author} {\bibinfo {author} {\bibfnamefont {D.~R.}\ \bibnamefont {Nelson}}\ and\ \bibinfo {author} {\bibfnamefont {B.~I.}\ \bibnamefont {Halperin}},\ }\bibfield  {title} {\enquote {\bibinfo {title} {Dislocation-mediated melting in two dimensions},}\ }\href {\doibase 10.1103/physrevb.19.2457} {\bibfield  {journal} {\bibinfo  {journal} {Physical Review B}\ }\textbf {\bibinfo {volume} {19}},\ \bibinfo {pages} {2457–2484} (\bibinfo {year} {1979})}\BibitemShut {NoStop}%
\bibitem [{\citenamefont {Strandburg}(1988)}]{Strandburg1988}%
  \BibitemOpen
  \bibfield  {author} {\bibinfo {author} {\bibfnamefont {K.~J.}\ \bibnamefont {Strandburg}},\ }\bibfield  {title} {\enquote {\bibinfo {title} {Two-dimensional melting},}\ }\href {\doibase 10.1103/revmodphys.60.161} {\bibfield  {journal} {\bibinfo  {journal} {Reviews of Modern Physics}\ }\textbf {\bibinfo {volume} {60}},\ \bibinfo {pages} {161–207} (\bibinfo {year} {1988})}\BibitemShut {NoStop}%
\bibitem [{\citenamefont {Bernard}\ and\ \citenamefont {Krauth}(2011)}]{Bernard2011}%
  \BibitemOpen
  \bibfield  {author} {\bibinfo {author} {\bibfnamefont {E.~P.}\ \bibnamefont {Bernard}}\ and\ \bibinfo {author} {\bibfnamefont {W.}~\bibnamefont {Krauth}},\ }\bibfield  {title} {\enquote {\bibinfo {title} {Two-step melting in two dimensions: First-order liquid-hexatic transition},}\ }\href {\doibase 10.1103/PhysRevLett.107.155704} {\bibfield  {journal} {\bibinfo  {journal} {Physical Review Letters}\ }\textbf {\bibinfo {volume} {107}},\ \bibinfo {pages} {155704} (\bibinfo {year} {2011})}\BibitemShut {NoStop}%
\bibitem [{\citenamefont {Kosterlitz}(2016)}]{Kosterlitz2016}%
  \BibitemOpen
  \bibfield  {author} {\bibinfo {author} {\bibfnamefont {J.~M.}\ \bibnamefont {Kosterlitz}},\ }\bibfield  {title} {\enquote {\bibinfo {title} {Kosterlitz–thouless physics: a review of key issues},}\ }\href {\doibase 10.1088/0034-4885/79/2/026001} {\bibfield  {journal} {\bibinfo  {journal} {Rep. Prog. Phys.}\ }\textbf {\bibinfo {volume} {79}},\ \bibinfo {pages} {026001} (\bibinfo {year} {2016})}\BibitemShut {NoStop}%
\bibitem [{\citenamefont {Barros~de Oliveira}\ \emph {et~al.}(2006)\citenamefont {Barros~de Oliveira}, \citenamefont {Netz}, \citenamefont {Colla},\ and\ \citenamefont {Barbosa}}]{deOliveira2006}%
  \BibitemOpen
  \bibfield  {author} {\bibinfo {author} {\bibfnamefont {A.}~\bibnamefont {Barros~de Oliveira}}, \bibinfo {author} {\bibfnamefont {P.~A.}\ \bibnamefont {Netz}}, \bibinfo {author} {\bibfnamefont {T.}~\bibnamefont {Colla}}, \ and\ \bibinfo {author} {\bibfnamefont {M.~C.}\ \bibnamefont {Barbosa}},\ }\bibfield  {title} {\enquote {\bibinfo {title} {Thermodynamic and dynamic anomalies for a three-dimensional isotropic core-softened potential},}\ }\href {\doibase 10.1063/1.2168458} {\bibfield  {journal} {\bibinfo  {journal} {The Journal of Chemical Physics}\ }\textbf {\bibinfo {volume} {124}} (\bibinfo {year} {2006}),\ 10.1063/1.2168458}\BibitemShut {NoStop}%
\bibitem [{\citenamefont {de~Oliveira}\ \emph {et~al.}(2006)\citenamefont {de~Oliveira}, \citenamefont {Netz}, \citenamefont {Colla},\ and\ \citenamefont {Barbosa}}]{deOliveira2006b}%
  \BibitemOpen
  \bibfield  {author} {\bibinfo {author} {\bibfnamefont {A.~B.}\ \bibnamefont {de~Oliveira}}, \bibinfo {author} {\bibfnamefont {P.~A.}\ \bibnamefont {Netz}}, \bibinfo {author} {\bibfnamefont {T.}~\bibnamefont {Colla}}, \ and\ \bibinfo {author} {\bibfnamefont {M.~C.}\ \bibnamefont {Barbosa}},\ }\bibfield  {title} {\enquote {\bibinfo {title} {Structural anomalies for a three dimensional isotropic core-softened potential},}\ }\href {\doibase 10.1063/1.2357119} {\bibfield  {journal} {\bibinfo  {journal} {The Journal of Chemical Physics}\ }\textbf {\bibinfo {volume} {125}} (\bibinfo {year} {2006}),\ 10.1063/1.2357119}\BibitemShut {NoStop}%
\bibitem [{\citenamefont {Likos}(2001)}]{Likos2001}%
  \BibitemOpen
  \bibfield  {author} {\bibinfo {author} {\bibfnamefont {C.~N.}\ \bibnamefont {Likos}},\ }\bibfield  {title} {\enquote {\bibinfo {title} {Effective interactions in soft condensed matter physics},}\ }\href {\doibase 10.1016/S0370-1573(00)00141-1} {\bibfield  {journal} {\bibinfo  {journal} {Physics Reports}\ }\textbf {\bibinfo {volume} {348}},\ \bibinfo {pages} {267--439} (\bibinfo {year} {2001})}\BibitemShut {NoStop}%
\bibitem [{\citenamefont {Likos}(2006)}]{Likos06}%
  \BibitemOpen
  \bibfield  {author} {\bibinfo {author} {\bibfnamefont {C.~N.}\ \bibnamefont {Likos}},\ }\bibfield  {title} {\enquote {\bibinfo {title} {Soft matter with soft particles},}\ }\href {\doibase 10.1039/B601916C} {\bibfield  {journal} {\bibinfo  {journal} {Soft Matter}\ }\textbf {\bibinfo {volume} {2}},\ \bibinfo {pages} {478--498} (\bibinfo {year} {2006})}\BibitemShut {NoStop}%
\bibitem [{\citenamefont {Jagla}(1998)}]{jagla1998}%
  \BibitemOpen
  \bibfield  {author} {\bibinfo {author} {\bibfnamefont {E.~A.}\ \bibnamefont {Jagla}},\ }\bibfield  {title} {\enquote {\bibinfo {title} {Core-softened potentials and the anomalous properties of water},}\ }\href {\doibase 10.1103/PhysRevE.58.1478} {\bibfield  {journal} {\bibinfo  {journal} {Phys. Rev. E}\ }\textbf {\bibinfo {volume} {58}},\ \bibinfo {pages} {1478--1486} (\bibinfo {year} {1998})}\BibitemShut {NoStop}%
\bibitem [{\citenamefont {Jagla}(1999)}]{Jagla1999a}%
  \BibitemOpen
  \bibfield  {author} {\bibinfo {author} {\bibfnamefont {E.~A.}\ \bibnamefont {Jagla}},\ }\bibfield  {title} {\enquote {\bibinfo {title} {{Minimum energy configurations of repelling particles in two dimensions}},}\ }\href@noop {} {\bibfield  {journal} {\bibinfo  {journal} {Journal of Chemical Physics}\ }\textbf {\bibinfo {volume} {110}},\ \bibinfo {pages} {451--456} (\bibinfo {year} {1999})}\BibitemShut {NoStop}%
\bibitem [{\citenamefont {Camp}(2006)}]{Camp2006}%
  \BibitemOpen
  \bibfield  {author} {\bibinfo {author} {\bibfnamefont {P.~J.}\ \bibnamefont {Camp}},\ }\bibfield  {title} {\enquote {\bibinfo {title} {Structure and dynamics in a monolayer of core-softened particles},}\ }\href {\doibase 10.1016/j.molliq.2006.03.003} {\bibfield  {journal} {\bibinfo  {journal} {Journal of Molecular Liquids}\ }\textbf {\bibinfo {volume} {127}},\ \bibinfo {pages} {10–13} (\bibinfo {year} {2006})}\BibitemShut {NoStop}%
\bibitem [{\citenamefont {Pattabhiraman}\ and\ \citenamefont {Dijkstra}(2017)}]{Pattabhiraman2017}%
  \BibitemOpen
  \bibfield  {author} {\bibinfo {author} {\bibfnamefont {H.}~\bibnamefont {Pattabhiraman}}\ and\ \bibinfo {author} {\bibfnamefont {M.}~\bibnamefont {Dijkstra}},\ }\bibfield  {title} {\enquote {\bibinfo {title} {On the formation of stripe, sigma, and honeycomb phases in a core–corona system},}\ }\href {\doibase 10.1039/c7sm00254h} {\bibfield  {journal} {\bibinfo  {journal} {Soft Matter}\ }\textbf {\bibinfo {volume} {13}},\ \bibinfo {pages} {4418–4432} (\bibinfo {year} {2017})}\BibitemShut {NoStop}%
\bibitem [{\citenamefont {Jain}, \citenamefont {Errington},\ and\ \citenamefont {Truskett}(2014)}]{Jain2014}%
  \BibitemOpen
  \bibfield  {author} {\bibinfo {author} {\bibfnamefont {A.}~\bibnamefont {Jain}}, \bibinfo {author} {\bibfnamefont {J.~R.}\ \bibnamefont {Errington}}, \ and\ \bibinfo {author} {\bibfnamefont {T.~M.}\ \bibnamefont {Truskett}},\ }\bibfield  {title} {\enquote {\bibinfo {title} {Dimensionality and design of isotropic interactions that stabilize honeycomb, square, simple cubic, and diamond lattices},}\ }\href {\doibase 10.1103/physrevx.4.031049} {\bibfield  {journal} {\bibinfo  {journal} {Physical Review X}\ }\textbf {\bibinfo {volume} {4}} (\bibinfo {year} {2014}),\ 10.1103/physrevx.4.031049}\BibitemShut {NoStop}%
\bibitem [{\citenamefont {Ryzhov}\ \emph {et~al.}(2023)\citenamefont {Ryzhov}, \citenamefont {Gaiduk}, \citenamefont {Tareeva}, \citenamefont {Fomin},\ and\ \citenamefont {Tsiok}}]{Ryzhov2023}%
  \BibitemOpen
  \bibfield  {author} {\bibinfo {author} {\bibfnamefont {V.~N.}\ \bibnamefont {Ryzhov}}, \bibinfo {author} {\bibfnamefont {E.~A.}\ \bibnamefont {Gaiduk}}, \bibinfo {author} {\bibfnamefont {E.~E.}\ \bibnamefont {Tareeva}}, \bibinfo {author} {\bibfnamefont {Y.~D.}\ \bibnamefont {Fomin}}, \ and\ \bibinfo {author} {\bibfnamefont {E.~N.}\ \bibnamefont {Tsiok}},\ }\bibfield  {title} {\enquote {\bibinfo {title} {Melting scenarios of two-dimensional systems: Possibilities of computer simulation},}\ }\href {\doibase 10.1134/s1063776123070129} {\bibfield  {journal} {\bibinfo  {journal} {Journal of Experimental and Theoretical Physics}\ }\textbf {\bibinfo {volume} {137}},\ \bibinfo {pages} {125–150} (\bibinfo {year} {2023})}\BibitemShut {NoStop}%
\bibitem [{\citenamefont {Fomin}, \citenamefont {Tsiok},\ and\ \citenamefont {Ryzhov}(2019)}]{Fomin2019b}%
  \BibitemOpen
  \bibfield  {author} {\bibinfo {author} {\bibfnamefont {Y.}~\bibnamefont {Fomin}}, \bibinfo {author} {\bibfnamefont {E.}~\bibnamefont {Tsiok}}, \ and\ \bibinfo {author} {\bibfnamefont {V.}~\bibnamefont {Ryzhov}},\ }\bibfield  {title} {\enquote {\bibinfo {title} {The stripe phase of two-dimensional core-softened systems: Structure recognition},}\ }\href {\doibase 10.1016/j.physa.2019.121401} {\bibfield  {journal} {\bibinfo  {journal} {Physica A: Statistical Mechanics and its Applications}\ }\textbf {\bibinfo {volume} {527}},\ \bibinfo {pages} {121401} (\bibinfo {year} {2019})}\BibitemShut {NoStop}%
\bibitem [{\citenamefont {Olson~Reichhardt}, \citenamefont {Reichhardt},\ and\ \citenamefont {Bishop}(2010)}]{OlsonReichhardt2010}%
  \BibitemOpen
  \bibfield  {author} {\bibinfo {author} {\bibfnamefont {C.~J.}\ \bibnamefont {Olson~Reichhardt}}, \bibinfo {author} {\bibfnamefont {C.}~\bibnamefont {Reichhardt}}, \ and\ \bibinfo {author} {\bibfnamefont {A.~R.}\ \bibnamefont {Bishop}},\ }\bibfield  {title} {\enquote {\bibinfo {title} {Structural transitions, melting, and intermediate phases for stripe- and clump-forming systems},}\ }\href {\doibase 10.1103/physreve.82.041502} {\bibfield  {journal} {\bibinfo  {journal} {Physical Review E}\ }\textbf {\bibinfo {volume} {82}} (\bibinfo {year} {2010}),\ 10.1103/physreve.82.041502}\BibitemShut {NoStop}%
\bibitem [{\citenamefont {Prestipino}, \citenamefont {Saija},\ and\ \citenamefont {Giaquinta}(2012)}]{Prestipino2012}%
  \BibitemOpen
  \bibfield  {author} {\bibinfo {author} {\bibfnamefont {S.}~\bibnamefont {Prestipino}}, \bibinfo {author} {\bibfnamefont {F.}~\bibnamefont {Saija}}, \ and\ \bibinfo {author} {\bibfnamefont {P.~V.}\ \bibnamefont {Giaquinta}},\ }\bibfield  {title} {\enquote {\bibinfo {title} {{Hexatic phase and water-like anomalies in a two-dimensional fluid of particles with a weakly softened core}},}\ }\href@noop {} {\bibfield  {journal} {\bibinfo  {journal} {Journal of Chemical Physics}\ }\textbf {\bibinfo {volume} {137}},\ \bibinfo {pages} {104503} (\bibinfo {year} {2012})}\BibitemShut {NoStop}%
\bibitem [{\citenamefont {Fomin}(2021)}]{Fomin2021}%
  \BibitemOpen
  \bibfield  {author} {\bibinfo {author} {\bibfnamefont {Y.}~\bibnamefont {Fomin}},\ }\bibfield  {title} {\enquote {\bibinfo {title} {The phase diagram of a two-dimensional core-softened system with purely repulsive monotonic potential},}\ }\href {\doibase 10.1016/j.physa.2020.125519} {\bibfield  {journal} {\bibinfo  {journal} {Physica A: Statistical Mechanics and its Applications}\ }\textbf {\bibinfo {volume} {565}},\ \bibinfo {pages} {125519} (\bibinfo {year} {2021})}\BibitemShut {NoStop}%
\bibitem [{\citenamefont {Bordin}\ and\ \citenamefont {Barbosa}(2018)}]{Bordin2018}%
  \BibitemOpen
  \bibfield  {author} {\bibinfo {author} {\bibfnamefont {J.~R.}\ \bibnamefont {Bordin}}\ and\ \bibinfo {author} {\bibfnamefont {M.~C.}\ \bibnamefont {Barbosa}},\ }\bibfield  {title} {\enquote {\bibinfo {title} {Waterlike anomalies in a two-dimensional core-softened potential},}\ }\href {\doibase 10.1103/physreve.97.022604} {\bibfield  {journal} {\bibinfo  {journal} {Physical Review E}\ }\textbf {\bibinfo {volume} {97}} (\bibinfo {year} {2018}),\ 10.1103/physreve.97.022604}\BibitemShut {NoStop}%
\bibitem [{\citenamefont {Wassermair}\ \emph {et~al.}(2024)\citenamefont {Wassermair}, \citenamefont {Kahl}, \citenamefont {Roth},\ and\ \citenamefont {Archer}}]{Wassermair2024}%
  \BibitemOpen
  \bibfield  {author} {\bibinfo {author} {\bibfnamefont {M.}~\bibnamefont {Wassermair}}, \bibinfo {author} {\bibfnamefont {G.}~\bibnamefont {Kahl}}, \bibinfo {author} {\bibfnamefont {R.}~\bibnamefont {Roth}}, \ and\ \bibinfo {author} {\bibfnamefont {A.~J.}\ \bibnamefont {Archer}},\ }\bibfield  {title} {\enquote {\bibinfo {title} {Fingerprints of ordered self-assembled structures in the liquid phase of a hard-core, square-shoulder system},}\ }\href {\doibase 10.1063/5.0226954} {\bibfield  {journal} {\bibinfo  {journal} {The Journal of Chemical Physics}\ }\textbf {\bibinfo {volume} {161}} (\bibinfo {year} {2024}),\ 10.1063/5.0226954}\BibitemShut {NoStop}%
\bibitem [{\citenamefont {Kapfer}\ and\ \citenamefont {Krauth}(2015)}]{Kapfer2015}%
  \BibitemOpen
  \bibfield  {author} {\bibinfo {author} {\bibfnamefont {S.~C.}\ \bibnamefont {Kapfer}}\ and\ \bibinfo {author} {\bibfnamefont {W.}~\bibnamefont {Krauth}},\ }\bibfield  {title} {\enquote {\bibinfo {title} {Two-dimensional melting: From liquid-hexatic coexistence to continuous transitions},}\ }\href {\doibase 10.1103/physrevlett.114.035702} {\bibfield  {journal} {\bibinfo  {journal} {Physical Review Letters}\ }\textbf {\bibinfo {volume} {114}} (\bibinfo {year} {2015}),\ 10.1103/physrevlett.114.035702}\BibitemShut {NoStop}%
\bibitem [{\citenamefont {Chen}, \citenamefont {Kaplan},\ and\ \citenamefont {Mostoller}(1995)}]{Chen1995}%
  \BibitemOpen
  \bibfield  {author} {\bibinfo {author} {\bibfnamefont {K.}~\bibnamefont {Chen}}, \bibinfo {author} {\bibfnamefont {T.}~\bibnamefont {Kaplan}}, \ and\ \bibinfo {author} {\bibfnamefont {M.}~\bibnamefont {Mostoller}},\ }\bibfield  {title} {\enquote {\bibinfo {title} {Melting in two-dimensional lennard-jones systems: Observation of a metastable hexatic phase},}\ }\href {\doibase 10.1103/physrevlett.74.4019} {\bibfield  {journal} {\bibinfo  {journal} {Physical Review Letters}\ }\textbf {\bibinfo {volume} {74}},\ \bibinfo {pages} {4019–4022} (\bibinfo {year} {1995})}\BibitemShut {NoStop}%
\bibitem [{\citenamefont {Jaster}(1999)}]{Jaster1999}%
  \BibitemOpen
  \bibfield  {author} {\bibinfo {author} {\bibfnamefont {A.}~\bibnamefont {Jaster}},\ }\bibfield  {title} {\enquote {\bibinfo {title} {Computer simulations of the two-dimensional melting transition using hard disks},}\ }\href {\doibase 10.1103/physreve.59.2594} {\bibfield  {journal} {\bibinfo  {journal} {Physical Review E}\ }\textbf {\bibinfo {volume} {59}},\ \bibinfo {pages} {2594–2602} (\bibinfo {year} {1999})}\BibitemShut {NoStop}%
\bibitem [{\citenamefont {Marcus}\ and\ \citenamefont {Rice}(1996)}]{Marcus1996}%
  \BibitemOpen
  \bibfield  {author} {\bibinfo {author} {\bibfnamefont {A.~H.}\ \bibnamefont {Marcus}}\ and\ \bibinfo {author} {\bibfnamefont {S.~A.}\ \bibnamefont {Rice}},\ }\bibfield  {title} {\enquote {\bibinfo {title} {Observations of first-order liquid-to-hexatic and hexatic-to-solid phase transitions in a confined colloid suspension},}\ }\href {\doibase 10.1103/physrevlett.77.2577} {\bibfield  {journal} {\bibinfo  {journal} {Physical Review Letters}\ }\textbf {\bibinfo {volume} {77}},\ \bibinfo {pages} {2577–2580} (\bibinfo {year} {1996})}\BibitemShut {NoStop}%
\bibitem [{\citenamefont {Nosenko}\ \emph {et~al.}(2009)\citenamefont {Nosenko}, \citenamefont {Zhdanov}, \citenamefont {Ivlev}, \citenamefont {Knapek},\ and\ \citenamefont {Morfill}}]{Nosenko2009}%
  \BibitemOpen
  \bibfield  {author} {\bibinfo {author} {\bibfnamefont {V.}~\bibnamefont {Nosenko}}, \bibinfo {author} {\bibfnamefont {S.~K.}\ \bibnamefont {Zhdanov}}, \bibinfo {author} {\bibfnamefont {A.~V.}\ \bibnamefont {Ivlev}}, \bibinfo {author} {\bibfnamefont {C.~A.}\ \bibnamefont {Knapek}}, \ and\ \bibinfo {author} {\bibfnamefont {G.~E.}\ \bibnamefont {Morfill}},\ }\bibfield  {title} {\enquote {\bibinfo {title} {2d melting of plasma crystals: Equilibrium and nonequilibrium regimes},}\ }\href {\doibase 10.1103/physrevlett.103.015001} {\bibfield  {journal} {\bibinfo  {journal} {Physical Review Letters}\ }\textbf {\bibinfo {volume} {103}} (\bibinfo {year} {2009}),\ 10.1103/physrevlett.103.015001}\BibitemShut {NoStop}%
\bibitem [{\citenamefont {Guillamón}\ \emph {et~al.}(2009)\citenamefont {Guillamón}, \citenamefont {Suderow}, \citenamefont {Fernández-Pacheco}, \citenamefont {Sesé}, \citenamefont {Córdoba}, \citenamefont {De~Teresa}, \citenamefont {Ibarra},\ and\ \citenamefont {Vieira}}]{Guillamn2009}%
  \BibitemOpen
  \bibfield  {author} {\bibinfo {author} {\bibfnamefont {I.}~\bibnamefont {Guillamón}}, \bibinfo {author} {\bibfnamefont {H.}~\bibnamefont {Suderow}}, \bibinfo {author} {\bibfnamefont {A.}~\bibnamefont {Fernández-Pacheco}}, \bibinfo {author} {\bibfnamefont {J.}~\bibnamefont {Sesé}}, \bibinfo {author} {\bibfnamefont {R.}~\bibnamefont {Córdoba}}, \bibinfo {author} {\bibfnamefont {J.~M.}\ \bibnamefont {De~Teresa}}, \bibinfo {author} {\bibfnamefont {M.~R.}\ \bibnamefont {Ibarra}}, \ and\ \bibinfo {author} {\bibfnamefont {S.}~\bibnamefont {Vieira}},\ }\bibfield  {title} {\enquote {\bibinfo {title} {Direct observation of melting in a two-dimensional superconducting vortex lattice},}\ }\href {\doibase 10.1038/nphys1368} {\bibfield  {journal} {\bibinfo  {journal} {Nature Physics}\ }\textbf {\bibinfo {volume} {5}},\ \bibinfo {pages} {651–655} (\bibinfo {year} {2009})}\BibitemShut {NoStop}%
\bibitem [{\citenamefont {Horn}\ \emph {et~al.}(2013)\citenamefont {Horn}, \citenamefont {Deutschl\"{a}nder}, \citenamefont {L\"{o}wen}, \citenamefont {Maret},\ and\ \citenamefont {Keim}}]{Horn2013}%
  \BibitemOpen
  \bibfield  {author} {\bibinfo {author} {\bibfnamefont {T.}~\bibnamefont {Horn}}, \bibinfo {author} {\bibfnamefont {S.}~\bibnamefont {Deutschl\"{a}nder}}, \bibinfo {author} {\bibfnamefont {H.}~\bibnamefont {L\"{o}wen}}, \bibinfo {author} {\bibfnamefont {G.}~\bibnamefont {Maret}}, \ and\ \bibinfo {author} {\bibfnamefont {P.}~\bibnamefont {Keim}},\ }\bibfield  {title} {\enquote {\bibinfo {title} {Fluctuations of orientational order and clustering in a two-dimensional colloidal system under quenched disorder},}\ }\href {\doibase 10.1103/physreve.88.062305} {\bibfield  {journal} {\bibinfo  {journal} {Physical Review E}\ }\textbf {\bibinfo {volume} {88}} (\bibinfo {year} {2013}),\ 10.1103/physreve.88.062305}\BibitemShut {NoStop}%
\bibitem [{\citenamefont {Deutschl\"{a}nder}\ \emph {et~al.}(2013)\citenamefont {Deutschl\"{a}nder}, \citenamefont {Horn}, \citenamefont {L\"{o}wen}, \citenamefont {Maret},\ and\ \citenamefont {Keim}}]{Deutschlander2013}%
  \BibitemOpen
  \bibfield  {author} {\bibinfo {author} {\bibfnamefont {S.}~\bibnamefont {Deutschl\"{a}nder}}, \bibinfo {author} {\bibfnamefont {T.}~\bibnamefont {Horn}}, \bibinfo {author} {\bibfnamefont {H.}~\bibnamefont {L\"{o}wen}}, \bibinfo {author} {\bibfnamefont {G.}~\bibnamefont {Maret}}, \ and\ \bibinfo {author} {\bibfnamefont {P.}~\bibnamefont {Keim}},\ }\bibfield  {title} {\enquote {\bibinfo {title} {Two-dimensional melting under quenched disorder},}\ }\href {\doibase 10.1103/physrevlett.111.098301} {\bibfield  {journal} {\bibinfo  {journal} {Physical Review Letters}\ }\textbf {\bibinfo {volume} {111}} (\bibinfo {year} {2013}),\ 10.1103/physrevlett.111.098301}\BibitemShut {NoStop}%
\bibitem [{\citenamefont {Hua}\ and\ \citenamefont {Han}(2024)}]{Hua2024}%
  \BibitemOpen
  \bibfield  {author} {\bibinfo {author} {\bibfnamefont {P.}~\bibnamefont {Hua}}\ and\ \bibinfo {author} {\bibfnamefont {Y.}~\bibnamefont {Han}},\ }\bibfield  {title} {\enquote {\bibinfo {title} {Searching for various melting scenarios of 2d crystals},}\ }\href {\doibase 10.1016/j.matt.2023.12.013} {\bibfield  {journal} {\bibinfo  {journal} {Matter}\ }\textbf {\bibinfo {volume} {7}},\ \bibinfo {pages} {19–22} (\bibinfo {year} {2024})}\BibitemShut {NoStop}%
\bibitem [{\citenamefont {Massana-Cid}\ \emph {et~al.}(2024)\citenamefont {Massana-Cid}, \citenamefont {Maggi}, \citenamefont {Gnan}, \citenamefont {Frangipane},\ and\ \citenamefont {Di~Leonardo}}]{MassanaCid2024}%
  \BibitemOpen
  \bibfield  {author} {\bibinfo {author} {\bibfnamefont {H.}~\bibnamefont {Massana-Cid}}, \bibinfo {author} {\bibfnamefont {C.}~\bibnamefont {Maggi}}, \bibinfo {author} {\bibfnamefont {N.}~\bibnamefont {Gnan}}, \bibinfo {author} {\bibfnamefont {G.}~\bibnamefont {Frangipane}}, \ and\ \bibinfo {author} {\bibfnamefont {R.}~\bibnamefont {Di~Leonardo}},\ }\bibfield  {title} {\enquote {\bibinfo {title} {Multiple temperatures and melting of a colloidal active crystal},}\ }\href {\doibase 10.1038/s41467-024-50937-2} {\bibfield  {journal} {\bibinfo  {journal} {Nature Communications}\ }\textbf {\bibinfo {volume} {15}} (\bibinfo {year} {2024}),\ 10.1038/s41467-024-50937-2}\BibitemShut {NoStop}%
\bibitem [{\citenamefont {Karthika}, \citenamefont {Radhakrishnan},\ and\ \citenamefont {Kalaichelvi}(2016)}]{Karthika2016}%
  \BibitemOpen
  \bibfield  {author} {\bibinfo {author} {\bibfnamefont {S.}~\bibnamefont {Karthika}}, \bibinfo {author} {\bibfnamefont {T.~K.}\ \bibnamefont {Radhakrishnan}}, \ and\ \bibinfo {author} {\bibfnamefont {P.}~\bibnamefont {Kalaichelvi}},\ }\bibfield  {title} {\enquote {\bibinfo {title} {A review of classical and nonclassical nucleation theories},}\ }\href {\doibase 10.1021/acs.cgd.6b00794} {\bibfield  {journal} {\bibinfo  {journal} {Crystal Growth \& Design}\ }\textbf {\bibinfo {volume} {16}},\ \bibinfo {pages} {6663–6681} (\bibinfo {year} {2016})}\BibitemShut {NoStop}%
\bibitem [{\citenamefont {Le~Cunuder}\ \emph {et~al.}(2019)\citenamefont {Le~Cunuder}, \citenamefont {Fr\'erot}, \citenamefont {Ortiz-Ambriz},\ and\ \citenamefont {Tierno}}]{Cunuder2019}%
  \BibitemOpen
  \bibfield  {author} {\bibinfo {author} {\bibfnamefont {A.}~\bibnamefont {Le~Cunuder}}, \bibinfo {author} {\bibfnamefont {I.}~\bibnamefont {Fr\'erot}}, \bibinfo {author} {\bibfnamefont {A.}~\bibnamefont {Ortiz-Ambriz}}, \ and\ \bibinfo {author} {\bibfnamefont {P.}~\bibnamefont {Tierno}},\ }\bibfield  {title} {\enquote {\bibinfo {title} {Competing orders in colloidal kagome ice: Importance of the in-trap motion of the particles},}\ }\href {\doibase 10.1103/PhysRevB.99.140405} {\bibfield  {journal} {\bibinfo  {journal} {Phys. Rev. B}\ }\textbf {\bibinfo {volume} {99}},\ \bibinfo {pages} {140405} (\bibinfo {year} {2019})}\BibitemShut {NoStop}%
\bibitem [{\citenamefont {Lib\'al}\ \emph {et~al.}(2018)\citenamefont {Lib\'al}, \citenamefont {Nisoli}, \citenamefont {Reichhardt},\ and\ \citenamefont {Reichhardt}}]{Libal2018}%
  \BibitemOpen
  \bibfield  {author} {\bibinfo {author} {\bibfnamefont {A.}~\bibnamefont {Lib\'al}}, \bibinfo {author} {\bibfnamefont {C.}~\bibnamefont {Nisoli}}, \bibinfo {author} {\bibfnamefont {C.~J.~O.}\ \bibnamefont {Reichhardt}}, \ and\ \bibinfo {author} {\bibfnamefont {C.}~\bibnamefont {Reichhardt}},\ }\bibfield  {title} {\enquote {\bibinfo {title} {Inner phases of colloidal hexagonal spin ice},}\ }\href {\doibase 10.1103/PhysRevLett.120.027204} {\bibfield  {journal} {\bibinfo  {journal} {Phys. Rev. Lett.}\ }\textbf {\bibinfo {volume} {120}},\ \bibinfo {pages} {027204} (\bibinfo {year} {2018})}\BibitemShut {NoStop}%
\bibitem [{\citenamefont {Nowack}\ and\ \citenamefont {Rice}(2019)}]{Nowack2019}%
  \BibitemOpen
  \bibfield  {author} {\bibinfo {author} {\bibfnamefont {L.}~\bibnamefont {Nowack}}\ and\ \bibinfo {author} {\bibfnamefont {S.~A.}\ \bibnamefont {Rice}},\ }\bibfield  {title} {\enquote {\bibinfo {title} {Sequential phase transitions and transient structured fluctuations in two-dimensional systems with a high-density kagome lattice phase},}\ }\href {\doibase 10.1063/1.5130558} {\bibfield  {journal} {\bibinfo  {journal} {The Journal of Chemical Physics}\ }\textbf {\bibinfo {volume} {151}} (\bibinfo {year} {2019}),\ 10.1063/1.5130558}\BibitemShut {NoStop}%
\bibitem [{\citenamefont {Anderson}\ \emph {et~al.}(2017)\citenamefont {Anderson}, \citenamefont {Antonaglia}, \citenamefont {Millan}, \citenamefont {Engel},\ and\ \citenamefont {Glotzer}}]{Anderson2017}%
  \BibitemOpen
  \bibfield  {author} {\bibinfo {author} {\bibfnamefont {J.~A.}\ \bibnamefont {Anderson}}, \bibinfo {author} {\bibfnamefont {J.}~\bibnamefont {Antonaglia}}, \bibinfo {author} {\bibfnamefont {J.~A.}\ \bibnamefont {Millan}}, \bibinfo {author} {\bibfnamefont {M.}~\bibnamefont {Engel}}, \ and\ \bibinfo {author} {\bibfnamefont {S.~C.}\ \bibnamefont {Glotzer}},\ }\bibfield  {title} {\enquote {\bibinfo {title} {Shape and symmetry determine two-dimensional melting transitions of hard regular polygons},}\ }\href {\doibase 10.1103/physrevx.7.021001} {\bibfield  {journal} {\bibinfo  {journal} {Physical Review X}\ }\textbf {\bibinfo {volume} {7}} (\bibinfo {year} {2017}),\ 10.1103/physrevx.7.021001}\BibitemShut {NoStop}%
\bibitem [{\citenamefont {Charbonneau}\ and\ \citenamefont {coauthors}(2023)}]{Charbonneau2024}%
  \BibitemOpen
  \bibfield  {author} {\bibinfo {author} {\bibfnamefont {P.}~\bibnamefont {Charbonneau}}\ and\ \bibinfo {author} {\bibnamefont {coauthors}},\ }\bibfield  {title} {\enquote {\bibinfo {title} {The complex physics of 2d melting},}\ }\href {\doibase 10.1103/PhysRevE.108.024103} {\bibfield  {journal} {\bibinfo  {journal} {Phys. Rev. E}\ }\textbf {\bibinfo {volume} {108}},\ \bibinfo {pages} {024103} (\bibinfo {year} {2023})}\BibitemShut {NoStop}%
\bibitem [{\citenamefont {Royall}\ \emph {et~al.}(2024)\citenamefont {Royall}, \citenamefont {Charbonneau}, \citenamefont {Dijkstra}, \citenamefont {Russo}, \citenamefont {Smallenburg}, \citenamefont {Speck},\ and\ \citenamefont {Valeriani}}]{Royall2024}%
  \BibitemOpen
  \bibfield  {author} {\bibinfo {author} {\bibfnamefont {C.~P.}\ \bibnamefont {Royall}}, \bibinfo {author} {\bibfnamefont {P.}~\bibnamefont {Charbonneau}}, \bibinfo {author} {\bibfnamefont {M.}~\bibnamefont {Dijkstra}}, \bibinfo {author} {\bibfnamefont {J.}~\bibnamefont {Russo}}, \bibinfo {author} {\bibfnamefont {F.}~\bibnamefont {Smallenburg}}, \bibinfo {author} {\bibfnamefont {T.}~\bibnamefont {Speck}}, \ and\ \bibinfo {author} {\bibfnamefont {C.}~\bibnamefont {Valeriani}},\ }\bibfield  {title} {\enquote {\bibinfo {title} {Colloidal hard spheres: Triumphs, challenges, and mysteries},}\ }\href {\doibase 10.1103/RevModPhys.96.045003} {\bibfield  {journal} {\bibinfo  {journal} {Rev. Mod. Phys.}\ }\textbf {\bibinfo {volume} {96}},\ \bibinfo {pages} {045003} (\bibinfo {year} {2024})}\BibitemShut {NoStop}%
\bibitem [{\citenamefont {Tsiok}\ \emph {et~al.}(2015)\citenamefont {Tsiok}, \citenamefont {Dudalov}, \citenamefont {Fomin},\ and\ \citenamefont {Ryzhov}}]{Tsiok2015}%
  \BibitemOpen
  \bibfield  {author} {\bibinfo {author} {\bibfnamefont {E.~N.}\ \bibnamefont {Tsiok}}, \bibinfo {author} {\bibfnamefont {D.~E.}\ \bibnamefont {Dudalov}}, \bibinfo {author} {\bibfnamefont {Y.~D.}\ \bibnamefont {Fomin}}, \ and\ \bibinfo {author} {\bibfnamefont {V.~N.}\ \bibnamefont {Ryzhov}},\ }\bibfield  {title} {\enquote {\bibinfo {title} {Random pinning changes the melting scenario of a two-dimensional core-softened potential system},}\ }\href {\doibase 10.1103/physreve.92.032110} {\bibfield  {journal} {\bibinfo  {journal} {Physical Review E}\ }\textbf {\bibinfo {volume} {92}} (\bibinfo {year} {2015}),\ 10.1103/physreve.92.032110}\BibitemShut {NoStop}%
\bibitem [{\citenamefont {Mendoza-Coto}\ \emph {et~al.}(2024)\citenamefont {Mendoza-Coto}, \citenamefont {Mattiello}, \citenamefont {Cenci}, \citenamefont {Defenu},\ and\ \citenamefont {Nicolao}}]{MendozaCoto2024}%
  \BibitemOpen
  \bibfield  {author} {\bibinfo {author} {\bibfnamefont {A.}~\bibnamefont {Mendoza-Coto}}, \bibinfo {author} {\bibfnamefont {V.}~\bibnamefont {Mattiello}}, \bibinfo {author} {\bibfnamefont {R.}~\bibnamefont {Cenci}}, \bibinfo {author} {\bibfnamefont {N.}~\bibnamefont {Defenu}}, \ and\ \bibinfo {author} {\bibfnamefont {L.}~\bibnamefont {Nicolao}},\ }\bibfield  {title} {\enquote {\bibinfo {title} {Melting of the two-dimensional solid phase in the gaussian core model},}\ }\href {\doibase 10.1103/physrevb.109.064101} {\bibfield  {journal} {\bibinfo  {journal} {Physical Review B}\ }\textbf {\bibinfo {volume} {109}} (\bibinfo {year} {2024}),\ 10.1103/physrevb.109.064101}\BibitemShut {NoStop}%
\bibitem [{\citenamefont {Stoycheva}\ and\ \citenamefont {Singer}(2000)}]{Stoycheva2000}%
  \BibitemOpen
  \bibfield  {author} {\bibinfo {author} {\bibfnamefont {A.~D.}\ \bibnamefont {Stoycheva}}\ and\ \bibinfo {author} {\bibfnamefont {S.~J.}\ \bibnamefont {Singer}},\ }\bibfield  {title} {\enquote {\bibinfo {title} {Stripe melting in a two-dimensional system with competing interactions},}\ }\href {\doibase 10.1103/physrevlett.84.4657} {\bibfield  {journal} {\bibinfo  {journal} {Physical Review Letters}\ }\textbf {\bibinfo {volume} {84}},\ \bibinfo {pages} {4657–4660} (\bibinfo {year} {2000})}\BibitemShut {NoStop}%
\bibitem [{\citenamefont {Cardoso}\ \emph {et~al.}(2021)\citenamefont {Cardoso}, \citenamefont {Hernandes}, \citenamefont {Nogueira},\ and\ \citenamefont {Bordin}}]{Cardoso2021}%
  \BibitemOpen
  \bibfield  {author} {\bibinfo {author} {\bibfnamefont {D.~S.}\ \bibnamefont {Cardoso}}, \bibinfo {author} {\bibfnamefont {V.~F.}\ \bibnamefont {Hernandes}}, \bibinfo {author} {\bibfnamefont {T.}~\bibnamefont {Nogueira}}, \ and\ \bibinfo {author} {\bibfnamefont {J.~R.}\ \bibnamefont {Bordin}},\ }\bibfield  {title} {\enquote {\bibinfo {title} {Structural behavior of a two length scale core-softened fluid in two dimensions},}\ }\href {\doibase 10.1016/j.physa.2020.125628} {\bibfield  {journal} {\bibinfo  {journal} {Physica A: Statistical Mechanics and its Applications}\ }\textbf {\bibinfo {volume} {566}},\ \bibinfo {pages} {125628} (\bibinfo {year} {2021})}\BibitemShut {NoStop}%
\bibitem [{\citenamefont {Nogueira}\ and\ \citenamefont {Bordin}(2022)}]{Nogueira2022}%
  \BibitemOpen
  \bibfield  {author} {\bibinfo {author} {\bibfnamefont {T.}~\bibnamefont {Nogueira}}\ and\ \bibinfo {author} {\bibfnamefont {J.~R.}\ \bibnamefont {Bordin}},\ }\bibfield  {title} {\enquote {\bibinfo {title} {Patterns in 2d core-softened systems: From sphere to dumbbell colloids},}\ }\href {\doibase 10.1016/j.physa.2022.128048} {\bibfield  {journal} {\bibinfo  {journal} {Physica A: Statistical Mechanics and its Applications}\ }\textbf {\bibinfo {volume} {605}},\ \bibinfo {pages} {128048} (\bibinfo {year} {2022})}\BibitemShut {NoStop}%
\bibitem [{\citenamefont {Nogueira}\ and\ \citenamefont {Bordin}(2023)}]{Nogueira2023}%
  \BibitemOpen
  \bibfield  {author} {\bibinfo {author} {\bibfnamefont {T.}~\bibnamefont {Nogueira}}\ and\ \bibinfo {author} {\bibfnamefont {J.~R.}\ \bibnamefont {Bordin}},\ }\bibfield  {title} {\enquote {\bibinfo {title} {Stripes polymorphism and water-like anomaly in hard core-soft corona dumbbells},}\ }\href {\doibase 10.1016/j.molliq.2023.123127} {\bibfield  {journal} {\bibinfo  {journal} {Journal of Molecular Liquids}\ }\textbf {\bibinfo {volume} {390}},\ \bibinfo {pages} {123127} (\bibinfo {year} {2023})}\BibitemShut {NoStop}%
\bibitem [{\citenamefont {Puccinelli}, \citenamefont {Ilha},\ and\ \citenamefont {Bordin}(2025)}]{Puccinelli25}%
  \BibitemOpen
  \bibfield  {author} {\bibinfo {author} {\bibfnamefont {T.}~\bibnamefont {Puccinelli}}, \bibinfo {author} {\bibfnamefont {A.~V.}\ \bibnamefont {Ilha}}, \ and\ \bibinfo {author} {\bibfnamefont {J.~R.}\ \bibnamefont {Bordin}},\ }\href {\doibase 10.48550/ARXIV.2507.13573} {\enquote {\bibinfo {title} {Competing length scales and symmetry frustration govern non-universal melting in 2d core-softened colloidal crystals},}\ } (\bibinfo {year} {2025})\BibitemShut {NoStop}%
\bibitem [{\citenamefont {Menath}\ \emph {et~al.}(2021)\citenamefont {Menath}, \citenamefont {Eatson}, \citenamefont {Brilmayer}, \citenamefont {Andrieu-Brunsen}, \citenamefont {Buzza},\ and\ \citenamefont {Vogel}}]{Menath2021}%
  \BibitemOpen
  \bibfield  {author} {\bibinfo {author} {\bibfnamefont {J.}~\bibnamefont {Menath}}, \bibinfo {author} {\bibfnamefont {J.}~\bibnamefont {Eatson}}, \bibinfo {author} {\bibfnamefont {R.}~\bibnamefont {Brilmayer}}, \bibinfo {author} {\bibfnamefont {A.}~\bibnamefont {Andrieu-Brunsen}}, \bibinfo {author} {\bibfnamefont {D.~M.~A.}\ \bibnamefont {Buzza}}, \ and\ \bibinfo {author} {\bibfnamefont {N.}~\bibnamefont {Vogel}},\ }\bibfield  {title} {\enquote {\bibinfo {title} {Defined core–shell particles as the key to complex interfacial self-assembly},}\ }\href {\doibase 10.1073/pnas.2113394118} {\bibfield  {journal} {\bibinfo  {journal} {Proceedings of the National Academy of Sciences}\ }\textbf {\bibinfo {volume} {118}} (\bibinfo {year} {2021}),\ 10.1073/pnas.2113394118}\BibitemShut {NoStop}%
\bibitem [{\citenamefont {Feller}\ and\ \citenamefont {Karg}(2022)}]{Feller2022}%
  \BibitemOpen
  \bibfield  {author} {\bibinfo {author} {\bibfnamefont {D.}~\bibnamefont {Feller}}\ and\ \bibinfo {author} {\bibfnamefont {M.}~\bibnamefont {Karg}},\ }\bibfield  {title} {\enquote {\bibinfo {title} {Fluid interface-assisted assembly of soft microgels: recent developments for structures beyond hexagonal packing},}\ }\href {\doibase 10.1039/d2sm00872f} {\bibfield  {journal} {\bibinfo  {journal} {Soft Matter}\ }\textbf {\bibinfo {volume} {18}},\ \bibinfo {pages} {6301–6312} (\bibinfo {year} {2022})}\BibitemShut {NoStop}%
\bibitem [{\citenamefont {Camerin}\ and\ \citenamefont {Zaccarelli}(2022)}]{Camerin2022}%
  \BibitemOpen
  \bibfield  {author} {\bibinfo {author} {\bibfnamefont {F.}~\bibnamefont {Camerin}}\ and\ \bibinfo {author} {\bibfnamefont {E.}~\bibnamefont {Zaccarelli}},\ }\bibfield  {title} {\enquote {\bibinfo {title} {Soft colloids for complex interfacial assemblies},}\ }\href {\doibase 10.1073/pnas.2122051119} {\bibfield  {journal} {\bibinfo  {journal} {Proceedings of the National Academy of Sciences}\ }\textbf {\bibinfo {volume} {119}} (\bibinfo {year} {2022}),\ 10.1073/pnas.2122051119}\BibitemShut {NoStop}%
\bibitem [{\citenamefont {Russo}\ and\ \citenamefont {Tanaka}(2018)}]{Russo2018}%
  \BibitemOpen
  \bibfield  {author} {\bibinfo {author} {\bibfnamefont {J.}~\bibnamefont {Russo}}\ and\ \bibinfo {author} {\bibfnamefont {H.}~\bibnamefont {Tanaka}},\ }\bibfield  {title} {\enquote {\bibinfo {title} {Entropic stabilization of open crystals from competition between local and global order},}\ }\href@noop {} {\bibfield  {journal} {\bibinfo  {journal} {Phys. Rev. X}\ }\textbf {\bibinfo {volume} {8}},\ \bibinfo {pages} {021040} (\bibinfo {year} {2018})}\BibitemShut {NoStop}%
\bibitem [{\citenamefont {Barros~de Oliveira}\ \emph {et~al.}(2010)\citenamefont {Barros~de Oliveira}, \citenamefont {Salcedo}, \citenamefont {Chakravarty},\ and\ \citenamefont {Barbosa}}]{BarrosdeOliveira2010}%
  \BibitemOpen
  \bibfield  {author} {\bibinfo {author} {\bibfnamefont {A.}~\bibnamefont {Barros~de Oliveira}}, \bibinfo {author} {\bibfnamefont {E.}~\bibnamefont {Salcedo}}, \bibinfo {author} {\bibfnamefont {C.}~\bibnamefont {Chakravarty}}, \ and\ \bibinfo {author} {\bibfnamefont {M.~C.}\ \bibnamefont {Barbosa}},\ }\bibfield  {title} {\enquote {\bibinfo {title} {Entropy, diffusivity and the energy landscape of a waterlike fluid},}\ }\href {\doibase 10.1063/1.3429254} {\bibfield  {journal} {\bibinfo  {journal} {The Journal of Chemical Physics}\ }\textbf {\bibinfo {volume} {132}} (\bibinfo {year} {2010}),\ 10.1063/1.3429254}\BibitemShut {NoStop}%
\bibitem [{\citenamefont {Thompson}\ \emph {et~al.}(2022)\citenamefont {Thompson}, \citenamefont {Aktulga}, \citenamefont {Berger}, \citenamefont {Bolintineanu}, \citenamefont {Brown}, \citenamefont {Crozier}, \citenamefont {in~'t Veld}, \citenamefont {Kohlmeyer}, \citenamefont {Moore}, \citenamefont {Nguyen}, \citenamefont {Shan}, \citenamefont {Stevens}, \citenamefont {Tranchida}, \citenamefont {Trott},\ and\ \citenamefont {Plimpton}}]{lammps}%
  \BibitemOpen
  \bibfield  {author} {\bibinfo {author} {\bibfnamefont {A.~P.}\ \bibnamefont {Thompson}}, \bibinfo {author} {\bibfnamefont {H.~M.}\ \bibnamefont {Aktulga}}, \bibinfo {author} {\bibfnamefont {R.}~\bibnamefont {Berger}}, \bibinfo {author} {\bibfnamefont {D.~S.}\ \bibnamefont {Bolintineanu}}, \bibinfo {author} {\bibfnamefont {W.~M.}\ \bibnamefont {Brown}}, \bibinfo {author} {\bibfnamefont {P.~S.}\ \bibnamefont {Crozier}}, \bibinfo {author} {\bibfnamefont {P.~J.}\ \bibnamefont {in~'t Veld}}, \bibinfo {author} {\bibfnamefont {A.}~\bibnamefont {Kohlmeyer}}, \bibinfo {author} {\bibfnamefont {S.~G.}\ \bibnamefont {Moore}}, \bibinfo {author} {\bibfnamefont {T.~D.}\ \bibnamefont {Nguyen}}, \bibinfo {author} {\bibfnamefont {R.}~\bibnamefont {Shan}}, \bibinfo {author} {\bibfnamefont {M.~J.}\ \bibnamefont {Stevens}}, \bibinfo {author} {\bibfnamefont {J.}~\bibnamefont {Tranchida}}, \bibinfo {author} {\bibfnamefont {C.}~\bibnamefont {Trott}}, \ and\ \bibinfo {author} {\bibfnamefont {S.~J.}\ \bibnamefont {Plimpton}},\
  }\bibfield  {title} {\enquote {\bibinfo {title} {{LAMMPS} - a flexible simulation tool for particle-based materials modeling at the atomic, meso, and continuum scales},}\ }\href {\doibase 10.1016/j.cpc.2021.108171} {\bibfield  {journal} {\bibinfo  {journal} {Comp. Phys. Comm.}\ }\textbf {\bibinfo {volume} {271}},\ \bibinfo {pages} {108171} (\bibinfo {year} {2022})}\BibitemShut {NoStop}%
\bibitem [{\citenamefont {Ramasubramani}\ \emph {et~al.}(2020)\citenamefont {Ramasubramani}, \citenamefont {Dice}, \citenamefont {Harper}, \citenamefont {Spellings}, \citenamefont {Anderson},\ and\ \citenamefont {Glotzer}}]{freud2020}%
  \BibitemOpen
  \bibfield  {author} {\bibinfo {author} {\bibfnamefont {V.}~\bibnamefont {Ramasubramani}}, \bibinfo {author} {\bibfnamefont {B.~D.}\ \bibnamefont {Dice}}, \bibinfo {author} {\bibfnamefont {E.~S.}\ \bibnamefont {Harper}}, \bibinfo {author} {\bibfnamefont {M.~P.}\ \bibnamefont {Spellings}}, \bibinfo {author} {\bibfnamefont {J.~A.}\ \bibnamefont {Anderson}}, \ and\ \bibinfo {author} {\bibfnamefont {S.~C.}\ \bibnamefont {Glotzer}},\ }\bibfield  {title} {\enquote {\bibinfo {title} {{freud: A software suite for high throughput analysis of particle simulation data}},}\ }\href@noop {} {\bibfield  {journal} {\bibinfo  {journal} {Computer Physics Communications}\ }\textbf {\bibinfo {volume} {254}},\ \bibinfo {pages} {107275} (\bibinfo {year} {2020})}\BibitemShut {NoStop}%
\bibitem [{\citenamefont {Errington}\ and\ \citenamefont {Debenedetti}(2001)}]{Errington2001}%
  \BibitemOpen
  \bibfield  {author} {\bibinfo {author} {\bibfnamefont {J.~R.}\ \bibnamefont {Errington}}\ and\ \bibinfo {author} {\bibfnamefont {P.~G.}\ \bibnamefont {Debenedetti}},\ }\bibfield  {title} {\enquote {\bibinfo {title} {{Relationship between structural order and the anomalies of liquid water}},}\ }\href@noop {} {\bibfield  {journal} {\bibinfo  {journal} {Nature}\ }\textbf {\bibinfo {volume} {409}},\ \bibinfo {pages} {318--321} (\bibinfo {year} {2001})}\BibitemShut {NoStop}%
\bibitem [{\citenamefont {Allen}, \citenamefont {Tildesley},\ and\ \citenamefont {Tildesley}(2017)}]{allen2017}%
  \BibitemOpen
  \bibfield  {author} {\bibinfo {author} {\bibfnamefont {M.}~\bibnamefont {Allen}}, \bibinfo {author} {\bibfnamefont {D.}~\bibnamefont {Tildesley}}, \ and\ \bibinfo {author} {\bibfnamefont {D.}~\bibnamefont {Tildesley}},\ }\href {https://books.google.com.br/books?id=nlExDwAAQBAJ} {\emph {\bibinfo {title} {Computer Simulation of Liquids}}}\ (\bibinfo  {publisher} {Oxford University Press},\ \bibinfo {year} {2017})\BibitemShut {NoStop}%
\bibitem [{\citenamefont {Malescio}\ and\ \citenamefont {Pellicane}(2003)}]{malescio2003}%
  \BibitemOpen
  \bibfield  {author} {\bibinfo {author} {\bibfnamefont {G.}~\bibnamefont {Malescio}}\ and\ \bibinfo {author} {\bibfnamefont {G.}~\bibnamefont {Pellicane}},\ }\bibfield  {title} {\enquote {\bibinfo {title} {Stripe phases from isotropic repulsive interactions},}\ }\href {\doibase 10.1038/nmat816} {\bibfield  {journal} {\bibinfo  {journal} {Nature Materials}\ }\textbf {\bibinfo {volume} {2}},\ \bibinfo {pages} {97--100} (\bibinfo {year} {2003})}\BibitemShut {NoStop}%
\bibitem [{\citenamefont {Han}\ \emph {et~al.}(2008)\citenamefont {Han}, \citenamefont {Shokef}, \citenamefont {Alsayed}, \citenamefont {Yunker}, \citenamefont {Lubensky},\ and\ \citenamefont {Yodh}}]{han2008}%
  \BibitemOpen
  \bibfield  {author} {\bibinfo {author} {\bibfnamefont {Y.}~\bibnamefont {Han}}, \bibinfo {author} {\bibfnamefont {Y.}~\bibnamefont {Shokef}}, \bibinfo {author} {\bibfnamefont {A.~M.}\ \bibnamefont {Alsayed}}, \bibinfo {author} {\bibfnamefont {P.}~\bibnamefont {Yunker}}, \bibinfo {author} {\bibfnamefont {T.~C.}\ \bibnamefont {Lubensky}}, \ and\ \bibinfo {author} {\bibfnamefont {A.~G.}\ \bibnamefont {Yodh}},\ }\bibfield  {title} {\enquote {\bibinfo {title} {Geometric frustration in buckled colloidal monolayers},}\ }\href {\doibase 10.1038/nature07441} {\bibfield  {journal} {\bibinfo  {journal} {Nature}\ }\textbf {\bibinfo {volume} {456}},\ \bibinfo {pages} {898--903} (\bibinfo {year} {2008})}\BibitemShut {NoStop}%
\end{thebibliography}%

\end{document}